\documentclass[12pt]{article}
\usepackage[a4paper, top=16mm, text={170mm, 248mm}, includehead, includefoot, hmarginratio=1:1, heightrounded]{geometry}
\usepackage{amsmath,amssymb,mathrsfs,amsthm,tikz,shuffle,paralist}
\usepackage{dsfont}
\usepackage{color}
\definecolor{darkred}{RGB}{173,34,48}

\usetikzlibrary{snakes}
\usetikzlibrary{calc}
\usetikzlibrary{decorations}

\usepackage[all]{xypic}
\usepackage{jheppub}
\usepackage{hyperref}

\newcommand{\dif}{\mathrm{d}} 
\newcommand{\R}{\mathbb{R}}
\DeclareMathOperator{\Li}{Li} 
\newcommand{\llangle}{\langle\!\langle}
\newcommand{\rrangle}{\rangle\!\rangle}

\title{The Wilson-loop $d \log$ representation for Feynman integrals}

\date{\today}
\author[a,b,c,d]{Song He}
\author[a,d]{Zhenjie Li} 
\author[e]{Yichao Tang}%
 \author[a,d]{Qinglin Yang}%
\affiliation[a]{CAS Key Laboratory of Theoretical Physics, Institute of Theoretical Physics, Chinese Academy of Sciences, Beijing 100190, China}
\affiliation[b]{
School of Fundamental Physics and Mathematical Sciences, Hangzhou Institute for Advanced Study, UCAS, Hangzhou 310024, China}
\affiliation[c]{ICTP-AP
International Centre for Theoretical Physics Asia-Pacific, Beijing/Hangzhou, China}
\affiliation[d]{School of Physical Sciences, University of Chinese Academy of Sciences, No.19A Yuquan Road, Beijing 100049, China}
\affiliation[e]{School of Physics, Peking University, Beijing 100871, China}
\emailAdd{songhe@itp.ac.cn}
\emailAdd{lizhenjie@itp.ac.cn}
\emailAdd{chaostang@pku.edu.cn}
\emailAdd{yangqinglin@itp.ac.cn}

\abstract{
We introduce and study the Wilson-loop ${\rm d}\log$ representation of certain Feynman integrals for scattering amplitudes in ${\cal N}=4$ SYM and beyond, which makes their evaluation completely straightforward. Such a representation was motivated by the dual Wilson loop picture, and it can also be derived by partial Feynman parametrization of loop integrals. We first introduce it for the simplest one-loop examples, the chiral pentagon in four dimensions and the three-mass-easy hexagon in six dimensions, which are represented by two- and three-fold ${\rm d}\log$ integrals that are nicely related to each other. For multi-loop examples, we write the $L$-loop generalized penta-ladders as $2(L{-}1)$-fold ${\rm d}\log$ integrals of some one-loop integral, so that once the latter is known, the integration can be performed in a systematic way. In particular, we write the eight-point penta-ladder as a $2L$-fold ${\rm d}\log$ integral whose symbol can be computed without performing any integration; we also obtain the last entries and the symbol alphabet of these integrals. Similarly we study the symbol of the seven-point double-penta-ladder, which is represented by a $2(L{-}1)$-fold integral of a hexagon; the latter can be written as a two-fold ${\rm d}\log$ integral plus a boundary term. We comment on the relation of our representation to differential equations and resumming the ladders by solving certain integral equations.}

\begin{document}

\maketitle

\section{Introduction}
Scattering amplitudes not only play a crucial role in connecting high-energy theory to experiments such as the Large Hadron Collider, but they are equally important in discovering new structures of Quantum Field Theory (QFT) and gravity. Tremendous progress has been made in the case of ${\cal N} = 4$ supersymmetric Yang-Mills theory (SYM), especially in the planar limit, but much remains to be understood. On the one hand, beautiful mathematical structures of the all-loop integrand~\cite{ArkaniHamed:2010kv, ArkaniHamed:2012nw,Arkani-Hamed:2013jha}) have been discovered, and on the other hand the (integrated) amplitudes have been computed to high loops, mostly for $n=6,7$~\cite{Dixon:2011pw,Dixon:2014xca,Dixon:2014iba,Drummond:2014ffa,Dixon:2015iva,Caron-Huot:2016owq,Chicherin:2017dob} but also for higher multiplicities~\cite{CaronHuot:2011ky,CaronHuot:2011kk,Zhang:2019vnm,He:2020vob}. 

There is a remarkable duality between maximally helicity-violating (MHV) scattering amplitudes and null polygonal Wilson loops (WL), discovered at both strong~\cite{Alday:2007hr,Alday:2007he} and weak couplings~\cite{Drummond:2007aua,Brandhuber:2007yx}, and much of the above progress has relied on this dual picture. Based on integrability~\cite{Beisert:2010jr} and operator product expansion (OPE) of WL~\cite{Alday:2010ku,Gaiotto:2010fk,Gaiotto:2011dt}, one can compute amplitudes at any value of the coupling around collinear limits~\cite{Basso:2013vsa}. The duality has been generalized to super-amplitudes and supersymmetric WL~\cite{Mason:2010yk,CaronHuot:2010ek}
, and they enjoy both superconformal and dual superconformal symmetries~\cite{Drummond:2006rz,Drummond:2008vq,Korchemsky:2010ut}, which close into the infinite-dimensional Yangian symmetry underpinning integrability~\cite{Drummond:2009fd}. Based on the super-WL picture, one can derive the powerful $\bar{Q}$ anomaly equations~\cite{CaronHuot:2011kk}, which allows us to compute multi-loop amplitudes directly from symmetries.

The extremely rich structure of all-loop scattering amplitudes/WL, at both the integrand and integrated-function level, has made ${\cal N}=4$ SYM a rather fruitful laboratory for studying Feynman loop integrals ({\it c.f.} \cite{Bourjaily:2018aeq,Henn:2018cdp,Herrmann:2019upk} and references therein), which is a subject of enormous interests and goes way beyond ${\cal N}=4$ SYM. In this paper, we propose to consider Feynman integrals, which naturally appear for amplitudes in ${\cal N}=4$ SYM and beyond, from the point of view of dual Wilson loops. We will see that for a large class of multi-loop Feynman integrals, rewriting them by exploiting WL picture largely simplifies and even trivializes their computation. Such ideas have been discussed in~\cite{He:2021}, and in particular our method is largely motivated by~\cite{CaronHuot:2010ek}. Alternatively, such WL representation of Feynman integrals could be obtained by partially Feynman parametrizing the loop integrals with respect to their massless corners.

In a companion paper~\cite{He:2021}, we show that both the one-loop chiral pentagon and generic two-loop double-pentagon integrals appear as Feynman diagrams of (super-) WL with insertion of fermions, integrated along $2$ and $4$ edges respectively. The dual picture motivates us to swap the order of integration: after first performing the loop integrals, we are left with simple line-integrals along the edges. We will generally refer to this representation of Feynman integrals as the Wilson-loop representation. The characteristic property of such a representation is that a pure integral of uniform transcendental weight $w$\footnote{In this paper, we mainly consider $L$-loop pure integrals in ${\cal N}=4$ SYM of weight $w=2L$, and a few other examples {\it e.g.} three-mass-easy hexagons in six dimensions of weight $w=3$.} is written as a $w$-fold integral of ${\rm d}\log$ forms, or a $w'$-fold ${\rm d}\log$ integral of pure, transcendental functions of weight $(w{-}w')$ (which by definition can also be written as $(w{-}w')$-fold ${\rm d}\log$ integrals). Once cast in such a form, we will call them Wilson-loop ${\rm d}\log$ integrals, even for cases where we do not explicitly relating them to actual WL diagrams.  

As the simplest example, we write the one-loop chiral pentagon as a two-fold ${\rm d}\log$ integral over two edges of the WL corresponding to its two massless corners; a similar example is the three-mass-easy hexagon in six dimensions, written as three-fold ${\rm d}\log$ integrals. There are certainly more one-loop integrals which can be written as WL ${\rm d}\log$ forms, such as four-mass boxes, but we will focus on cases with massless corners in this paper. Quite nicely, these representations naturally generalize to classes of multi-loop integrals, and in particular we will discuss the so-called generalized penta-ladder integrals with one-end being chiral pentagon and the other end generic. We find that an $L$-loop ladder can be written as a two-fold ${\rm d}\log$ integral of an $(L{-}1)$-loop one, and eventually we have a $2(L{-}1)$-fold integral of a one-loop integral, which combines the pentagon on the right end with an arbitrary one-loop object on the left end. The simplest case is the generic (eight-point) penta-ladder where the left end is a box, and the final one-loop integral is again a chiral pentagon. A slightly more nontrivial case is the double-penta-ladder, where the final one-loop integral is a hexagon. There are other multi-loop integrals with such ``terminal chiral-pentagon" sub-diagrams that we can write as ${\rm d}\log$ integrals of fewer-loop integrals.

The main advantage for considering WL ${\rm d}\log$ representations is that it essentially trivializes the evaluation of certain Feynman integrals. First of all, this representation makes properties such as dual conformal invariance (DCI) and transcendental weight completely manifest. More importantly, for Feynman integrals that evaluate to generalized polylogarithms of uniform weight, such a ${\rm d}\log$ integral representation captures its invariant contents and allows one to obtain the symbol~\cite{Goncharov:2010jf, Duhr:2011zq}, as long as entries to the ${\rm d}\log$ forms remain linear in the integration variables at all steps when the integrals are performed in order. 
Our representation guarantees this for a large class of integrals where the ${\rm d}\log$ forms are linear fractions of the integration variables. For more complicated cases, {\it e.g.} when the final one-loop integral for generalized penta-ladder contains square roots, the ${\rm d}\log$ form is no longer rational and we need to ``rationalize" the square roots to proceed. An important example is the double-pentagon integral (and higher-loop generalizations) with $n\geq 8$ legs, and we will discuss how to rationalize the square roots there in~\cite{He:2021}. 

For simplicity and to illustrate our point without distractions from ``rationalization", in the following we focus on cases where the ${\rm d}\log$ forms are rational, {\it e.g.} seven-point double-penta-ladders. For these integrals, we are essentially done once we arrive at the WL representation since it is linear in each variable with the obvious order of integration. There is still the non-trivial problem (depending on one's taste) of writing them as combinations of generalized polylogarithms~\footnote{In principle, this can always be accomplished using automated codes such as HyperInt~\cite{Panzer:2014caa} or polylogtools~\cite{Duhr:2019tlz}, though simplifying the results still requires some work.}. We will content ourselves in finding the {\it symbol} of these integrals, which is directly obtained from these ${\rm d}\log$ forms even without performing any integration. As discussed in~\cite{CaronHuot:2011kk}, there is a nice algorithm which we will review here for computing the symbol of ${\rm d}\log$ integrals in a purely algebraic way. Using this method, we present the symbol in a nice form for the eight-point penta-box and seven-point double-pentagon. We emphasize that the symbol of their higher-loop generalizations can also be obtained straightforwardly. Moreover, using the algorithm we can read off the last entries and and recursively prove other properties of the symbol for arbitrarily high loops, without even computing the full symbol.

For the Feynman integrals and amplitudes/WL we study, it is convenient to use momentum twistors~\cite{Hodges:2009hk}, which correspond to null rays of the dual spacetime and manifest the $SL(2,2)$ dual conformal symmetries~\cite{Drummond:2006rz,Drummond:2008vq,Korchemsky:2010ut}. For the polygonal Wilson loop, its vertices are given by $x_i$ with light-like edges $(x_{i{+}1}-x_i)^{\alpha \dot{\alpha}}=\lambda^\alpha_i \tilde\lambda^{\dot{\alpha}}_i$ for $i=1,2,\cdots, n$
, and similarly for the Grassmann part $(\theta_{i{+}1}-\theta_i)^{\alpha I}=\lambda_i^\alpha \eta_i^I$. The (supersymmetric) momentum twistors (one for each edge) are defined as
\[
 \mathcal{Z}_{i}=(Z_{i}^{a}\vert \chi_{i}^{A}):=(\lambda_{i}^{\alpha},x_{i}^{\alpha\dot{\alpha}}\lambda_{i\alpha}\vert\theta_{i}^{\alpha A}\lambda_{i\alpha}).
 \]
Any point in the dual spacetime corresponds to a line in twistor space, {\it e.g.} the vertex $x_i\sim (i{-}1i)$ corresponds to a line determined by two twistors $Z_{i{-}1}, Z_i$ (or the edges $i{-}1$ and $i$). Each loop momentum is represented by a dual (insertion) point, $x_\ell \sim \ell:=(A B)$, which is also represented by any two twistors $A,B$ on the line. The basic $SL(4)$ invariant is the four-bracket of four momentum twistors $\langle ijkl\rangle:=\epsilon_{a,b,c,d} Z_i^a Z_j^b Z_k^c Z_l^d $. The usual squared-distance of two dual points is proportional to the four-bracket using the corresponding two bi-twistors, {\it e.g.}  $(x_i-x_j)^2=\frac{\langle i{-}1ij{-}1j\rangle}{\langle i{-}1i\rangle \langle j{-}1j\rangle }$, and similarly $\langle \ell\,i{-}1 i\rangle:=\langle A B i{-}1 i\rangle=(x_\ell-x_i)^2 \langle A B \rangle \langle i{-}1 i\rangle$.\footnote{The two-brackets are the usual for spinors, which can be written as $\langle i j\rangle:=\langle i j I_{\infty}\rangle$ where $I_\infty$ denotes the infinity bi-twistor.}

In addition, we will be using generalized (Goncharov) polylogarithms and the symbol extensively, which we review here. Throughout the paper, our integrals evaluate to linear combinations of Goncharov polylogarithms of uniform weight $w$, defined by the $w$-fold iterated integrals~\cite{goncharov2005galois}
\begin{equation}
    G(a_{1},\ldots,a_{w};z):=\int_{0}^{z}\frac{\dif t}{t-a_{1}}\, G(a_{2},\ldots,a_{w};t), \label{Gpolylot} 
\end{equation} 
with the starting point $G(;z):=1$. It is straightforward to see that the differential of such a combination satisfies
\begin{equation}
    \dif G^{(w)}({\bf a}) = \sum_{i=1}^w G^{(w{-}1)}({\bf a}_i)\,\dif \log \frac{a_i-a_{i{-}1}}{a_i-a_{i{+}1}}
\end{equation}
where $G^{(w{-}1)}({\bf a}_i)$ are Goncharov polylogarithms of weight $w{-}1$ where we delete $a_i$ in the sequence and denote it as ${\bf a}_i$, accompanied by the differential $\dif \log \frac{a_i-a_{i{-}1}}{a_i-a_{i{+}1}}$ with boundary cases $a_0:=z$ and $a_{w{+}1}:=0$. Then, one can introduce a symbol map for polylogarithms by recursively defining
\begin{equation}
    \mathcal{S}(G^{(w)}({\bf a})) :=\sum_i \mathcal{S}(G^{(w-1)} ({\bf a}_i))\otimes \frac{a_i-a_{i{-}1}}{a_i-a_{i{+}1}}
\end{equation}
with $\mathcal{S}(\log a):= a$. We call these difference of adjacent $a_i$'s (including endpoints) generated in this way {\it symbol letters}, each tensor product of letters in the symbol a {\it{word}}, and the collection of all letters the {\it symbol alphabet}~\cite{Goncharov.A.B.:2009tja, Goncharov:2010jf}.

The rest of the paper is organized as follows. In section 2 we introduce the WL ${\rm d}\log$ representation using pedagogical one-loop examples of the chiral pentagon and the 6D three-mass-easy hexagon, and we review the algorithm for obtaining the symbol from ${\rm d}\log$ forms algebraically. We move to the WL ${\rm d}\log$ representation of the simplest multi-loop examples, penta-ladder integrals which depend on $3$ cross ratios (generic for $n=8$) in section 3. Starting from the chiral pentagon, we recursively write $L$-loop penta-ladder as a two-fold ${\rm d}\log$ integral of a $(L{-}1)$-loop penta-ladder with shifted kinematics, and recast the recursion relation in DCI form, naturally defining the odd-weight integrals along the way. We apply the algorithm to find $5$ last entries appearing in $3$ combinations for any $L$, and write down the symbol for the $L=2$ case explicitly. We then study generalized penta-ladder integrals in section 4, which differ from the previous case in that the starting point of the recursion can be any one-loop integral. We study the $n=7$ double-penta-ladder as the main example, which depends on $4$ cross ratios, and find $7$ last entries in $4$ combinations to all loops, as well as the explicit symbol for $L=2$; we also notice that the starting point, the $n=7$ hexagon, can be naturally written as a two-fold ${\rm d}\log$ integral, up to a boundary term which is a ${\rm d}\log$ integral of a $\log$ function. Finally in section 5, we comment on the relation between our WL representation and differential equations satisfied by ladder integrals~\cite{Drummond:2010cz}, and more importantly we resum these ladders by solving certain integral equations derived from our representation.

\section{Wilson-loop $\dif\log$ integrals: one-loop examples}
\label{sec.2}

In this section we present two one-loop examples of the Wilson-loop ${\rm d}\log$ representation, which are the chiral pentagon in four dimensions~\cite{ArkaniHamed:2010gh} and the three-mass-easy hexagon in six dimensions~\cite{DelDuca:2011wh, Spradlin:2011wp}. As we will see, the former can be nicely written as a two-fold ${\rm d}\log$ integral over two edges of the dual Wilson loops, and similarly the latter as a three-fold ${\rm d}\log$ integral. With four-dimensional kinematics, quite nicely the 6D three-mass-easy hexagon turns out to be a single ${\rm d}\log$ integral of the 4D chiral pentagon. The symbol for these weight-$2$ and $3$ integrals can be trivially obtained using the algorithm in \cite{CaronHuot:2011kk} which we review presently.

\subsection{Chiral pentagons}

Our prototypical example is the chiral pentagon integral defined as:\footnote{Throughout the paper, we absorb the standard prefactor $1/4 \pi^2$ into the measure $d^4 \ell$.}
\begin{equation}
    \Psi_1(i,j;I):=\begin{tikzpicture}[baseline={([yshift=-.5ex]current bounding box.center)},scale=0.18]
        \draw[black,thick] (0,0)--(5,0)--(6.55,4.76)--(2.50,7.69)--(-1.55,4.76)--cycle;
        \draw[decorate, decoration=snake, segment length=12pt, segment amplitude=2pt, black,thick] (6.55,4.76)--(-1.55,4.76);
        \draw[black,thick] (1.5,9.43)--(2.5,7.69)--(3.5,9.43);
        \filldraw[black] (2.5,9.19) circle [radius=2pt];
        \filldraw[black] (1.99,9.1) circle [radius=2pt];
        \filldraw[black] (3.01,9.1) circle [radius=2pt];
        \draw[black,thick] (-0.21,-1.99)--(0,0)--(-1.83,-0.81);
        \filldraw[black] (-0.88,-1.21) circle [radius=2pt];
        \filldraw[black] (-0.41,-1.44) circle [radius=2pt];
        \filldraw[black] (-1.24,-0.84) circle [radius=2pt];
        \draw[black,thick] (6.83,-0.81)--(5,0)--(5.21,-1.99);
        \filldraw[black] (5.88,-1.21) circle [radius=2pt];
        \filldraw[black] (6.24,-0.84) circle [radius=2pt];
        \filldraw[black] (5.41,-1.44) circle [radius=2pt];
        \draw[black,thick] (6.55,4.76)--(8.45,5.37);
        \draw[black,thick] (-1.55,4.76)--(-3.45,5.37);
        \filldraw[black] (-3.45,5.37) node[anchor=east] {{$i$}};
        \filldraw[black] (8.45,5.37) node[anchor=west] {{$j$}};
        \filldraw[black] (2.5,0) node[anchor=north] {{$I$}};
    \end{tikzpicture}=\int{\rm d}^4\ell\frac{\langle\ell\bar i\cap\bar j\rangle\langle Iij\rangle}{\llangle\ell i\rrangle\llangle\ell j\rrangle\langle\ell I\rangle},
\end{equation}
where for the two pairs of propagators adjacent to massless leg $i$ and $j$, we introduce the shorthand notation $\llangle\ell i\rrangle:=\langle\ell i{-}1i\rangle\langle\ell ii{+}1\rangle$ (and similarly for $j$); the last propagator depends on an arbitrary reference dual point $x_I\sim(I)$; and $\langle\ell\bar i\cap\bar j\rangle$ is determined by the line $\ell \sim (A B)$ and the intersection of two planes $\bar{i}, \bar{j}$: \[\langle\ell\bar{i}\cap\bar{j}\rangle=\langle Ai{-}1ii{+}1\rangle\langle Bj{-}1jj{+1}\rangle-\langle Bi{-}1ii{+}1\rangle\langle Aj{-}1jj{+1}\rangle\,.\]

Introducing line integrals in momentum twistor space using $\frac1{\llangle\ell i\rrangle}=\int_0^\infty\frac{{\rm d}\tau_X}{\langle\ell iX\rangle^2}$ with $X:=Z_{i-1}-\tau_XZ_{i+1}$ and a similar identity for $Y:=Z_{j-1}-\tau_YZ_{j+1}$, we see that
\begin{equation}
    \Psi_1(i,j;I)=\int{\rm d}^4\ell\int_0^\infty{\rm d}^2\tau\frac{\langle\ell\bar i\cap\bar j\rangle\langle Iij\rangle}{\langle\ell iX\rangle^2\langle\ell jY\rangle^2\langle\ell I\rangle}.
\end{equation}
Note that we have a natural interpretation of the $\tau_X$-integral: the insertion point $x\sim(iX)$ is integrated over the $i$-th edge of the Wilson loop, with endpoints $x_i\sim(i{-}1i)$ and $x_{i+1}\sim(ii{+}1)$ corresponding to $\tau_X=0$ and $\tau_X=\infty$, respectively. Similarly, the $\tau_Y$-integral is over the insertion of $y\sim(jY)$. Such integrals are familiar from the calculation of ${\bar Q}$ anomaly with fermion insertions~\cite{CaronHuot:2010ek, CaronHuot:2011kk}.

The crucial point is that the loop integral can now be performed directly, {\it e.g.} using Feynman parametrization, yielding a {\it rational} function. In terms of dual coordinates, the loop integral is (defining $w\sim(I)$) \[\int{\rm d}^4\ell\frac{(\ell-\bar z)^2(w-z)^2}{(\ell-x)^4(\ell-y)^4(\ell-w)^2}.\]
Notice that the Schubert problem solutions $z\sim(ij)$ and $\bar z\sim(\bar i\cap\bar j)$ are light-like separated from both $x$ and $y$. Introduce Feynman parameters $u_{1,2}$ and change to projective variables $\alpha_{1,2}:=u_{1,2}/(1-u_1-u_2)$:
\[\displaystyle\int_0^\infty{\rm d}\alpha_1{\rm d}\alpha_2\int{\rm d}^4\ell\frac{3\alpha_1\alpha_2(\boldsymbol\ell\cdot\bar{\boldsymbol z})(\boldsymbol w\cdot\boldsymbol z)}{[\boldsymbol\ell\cdot(\alpha_1\boldsymbol x+\alpha_2\boldsymbol y+\boldsymbol w)]^5},\]
where the boldface symbols are the embedded vectors in 6D \cite{Arkani-Hamed:2017ahv}. Define $\boldsymbol W:=\alpha_1\boldsymbol x+\alpha_2\boldsymbol y+\boldsymbol w$ so that $\bar{\boldsymbol z}\cdot\boldsymbol W=\bar{\boldsymbol z}\cdot\boldsymbol w$ by $(x-\bar z)^2=(y-\bar z)^2=0$.
\[\frac12\int_0^\infty{\rm d}\alpha_1{\rm d}\alpha_2\frac{\alpha_1\alpha_2(\boldsymbol w\cdot\boldsymbol z)(\bar{\boldsymbol z}\cdot\partial_{\boldsymbol W})}{(\boldsymbol W\cdot\boldsymbol W)^2}=\frac{(w-z)^2(w-\bar z)^2}{(x-y)^2(x-w)^2(y-w)^2}.\]
 Converting the above expression back to momentum twistors, we finally obtain:
\begin{equation}\label{startri}
   \int {\rm d}^4\ell\frac{\langle\ell\bar i\cap\bar j\rangle\langle Iij\rangle}{\langle\ell iX\rangle^2\langle\ell jY\rangle^2\langle\ell I\rangle}=\frac{\langle I\bar i\cap\bar j\rangle\langle Iij\rangle}{\langle iXjY\rangle\langle iXI\rangle\langle jYI\rangle}.
\end{equation}
which is known as the star-triangle identity. Thus we see that the chiral pentagon integral can be written as a two-fold integral over WL edges $i$ and $j$, of a rational integrand:
\begin{equation}\label{pentintegrand}
    \Psi_1(i,j;I)=\int_0^\infty{\rm d}^2\tau\frac{\langle I\bar i\cap\bar j\rangle\langle Iij\rangle}{\langle iXjY\rangle\langle iXI\rangle\langle jYI\rangle}.
\end{equation}

The integrand can be put into ${\rm d}\log$ forms to manifest its transcendental weight. By partial-fractioning with respect to first $\tau_X$ then $\tau_Y$, and using the identity $\frac{{\rm d}\tau}{a\tau+b}=\frac1a{\rm d}\log(a\tau+b)$, the result is
\begin{equation}\label{pentdlog}
 \boxed{\Psi_1(i,j;I)=\int_{\R^2_{\geq 0}}{\rm d}\log\frac{\langle jYI\rangle}{\langle jYiI\cap\bar i\rangle}{\rm d}\log\frac{\langle iXjY\rangle}{\langle iXI\rangle},}
\end{equation}
where the two ${\rm d}\log$ forms are differentials with respect to $\tau_Y$ and $\tau_X$, respectively, and (if the bi-twistor $(I)\equiv(ZZ')$)\[\langle jYiI\cap\bar i\rangle=\langle jYiZ\rangle\langle Z'i{-}1ii{+}1\rangle-\langle jYiZ'\rangle\langle Zi{-}1ii{+}1\rangle.\] It is clear that we should perform the $\tau_X$ integral before $\tau_Y$, and at each step the ${\rm d}\log$ is linear in the $\tau$'s, which will be a general feature of our WL ${\rm d}\log$ integrals. The $\tau_X$ integral trivially gives the $\log$ of a cross ratio that depends on $Y$, and then it is straightforward to perform the $\tau_Y$-integral to obtain
\begin{equation}
    \Psi_1(i,j;I)=\log u\log v+\Li_2(1{-}u)+\Li_2(1{-}v)+\Li_2(1{-}w)-\Li_2(1{-}uw)-\Li_2(1{-}vw),
\end{equation}
where the three cross ratios are defined as
\begin{equation}\label{crpenta}
    u=\dfrac{\langle i{-}1iI\rangle\langle jj{+}1ii{+}1\rangle}{\langle i{-}1ijj{+}1\rangle\langle Iii{+}1\rangle},\ v=\dfrac{\langle jj{+}1I\rangle\langle i{-}1ij{-}1j\rangle}{\langle jj{+}1i{-}1i\rangle\langle Ij{-}1j\rangle},\ w=\dfrac{\langle i{-}1ijj{+}1\rangle\langle j{-}1jii{+}1\rangle}{\langle i{-}1ij{-}1j\rangle\langle jj{+}1ii{+}1\rangle}.
\end{equation}

We remark that it would be interesting to understand the possible geometric meaning of this ${\rm d}\log$ integral. In fact, all ${\rm d}\log$ integrals considered in this paper are generalizations of the Aomoto polylogarithms, where we integrate the canonical ${\rm d}\log$ form of a simplex (or polytope) over another simplex (or polytope)~\cite{Goncharov.A.B.:2009tja,Aomoto:1982}; as shown in~\cite{Arkani-Hamed:2017ahv}, the symbol of such a integral can be extracted purely geometrically. In our cases we have positive geometries that go beyond polytopes: for the chiral pentagon, the $2$-form can be interpreted as the canonical form of a ``curvy" triangle, integrated over a normal triangle (in projective space); alternatively we could have used variables such that we integrate the canonical form of a normal triangle but over a ``curvy" region. 

\subsection{Integration of ${\rm d}\log$ forms at the symbol level}\label{sec.2.2}
Now we proceed to review an automated algorithm \cite{CaronHuot:2011kk} (see also \cite{Henn:2018cdp}) for computing the symbol algebraically from ${\rm d}\log$ forms. From the definition of symbol, the computation for 1-d integral amounts to nothing but taking differential. Suppose we have an integral
\[
\int_a^b {\rm d}\log(t+c)\, (F(t)\otimes w(t)),
\]
where $F(t)\otimes w(t)$ is a integrable, linear reducible symbol in $t$, {\it i.e.} its entries are products of powers of linear polynomials in $t$, and $w(t)$ is the last entry. Since the differential only acts on the last entry of a symbol, the total differential of this integral is the sum of the following two parts:
\begin{compactenum}[\quad (1)]
\item the contribution from endpoints:
\[
    {\rm d}\log(t+c)(F(t)\otimes w(t))|_{t=a}^{t=b}=(F(t)\otimes w(t)\otimes (t+c))|_{t=a}^{t=b},
\]
\item contributions from the last entry: for a term where $w(t)$ is a constant,
\[
\left(\int_a^b {\rm d}\log(t+c)\, F(t)\right){\rm d}\log w=\left(\int_a^b {\rm d}\log(t+c)\, F(t)\right)\otimes w,
\]
and for a term where $w(t)=t+d$,
\[
\left(\int_a^b {\rm d}\log \frac{t+c}{t+d}\, F(t)\right){\rm d}\log (c-d)
=\left(\int_a^b {\rm d}\log \frac{t+c}{t+d}\, F(t)\right)\otimes (c-d).
\]
\end{compactenum}
Then we can recursively compute the symbol of the lower-weight integrals and obtain the symbol of the iterated integral of ${\rm d} \log$ forms. 

As a trivial example, we apply this algorithm to obtain the symbol of $\Psi_1$. Before doing so, we note that although the integrand in \eqref{pentintegrand} itself is dual conformally invariant (DCI), entries of the ${\rm d}\log$ forms in \eqref{pentdlog} are not. It is desirable to make the DCI property manifest for the ${\rm d}\log$ forms, which can be achieved by a simple change of variables. Note that the integral is invariant if we rescale $\tau_{X}\rightarrow C_{X}\tau_{X}$ and similarly for $\tau_Y$, and one such choice is $C_X=\frac{\langle i-1ijY\rangle}{\langle ii+1jY\rangle}$ and $C_Y=\frac{\langle j-1jI\rangle}{\langle jj+1 I\rangle}$, which makes the new $\tau$-variables DCI (In contrast, the original $\tau_X$ has twistor weight $\frac{Z_{i-1}}{Z_{i+1}}$). This simple change of variables makes the ${\rm d}\log$ entries manifestly DCI:

\begin{equation}\label{I5uvw}
 \Psi_1=\int{\rm d}\log\frac{\tau_Y+1}{v(1-uw)\tau_Y+(1-u)}{\rm d}\log\frac{\tau_X+1}{u(\tau_Y+vw)\tau_X+(\tau_Y+v)}.
 \end{equation}

We leave it to the reader as a simple exercise to apply the algorithm to \eqref{I5uvw}: it remains manifestly DCI in intermediate steps, and the result for the symbol is
\begin{equation}\label{I5symbol}
\begin{split}
    \mathcal{S}(\Psi_1(u,v,w))&= u\otimes v+ v\otimes u-u\otimes(1-u)-v\otimes(1-v)\\
    &- w\otimes(1-w)+ (uw)\otimes (1-uw)+ (vw)\otimes(1-vw).
\end{split}
\end{equation}
This algorithm will be used for all our computations of the symbol in the remaining of the paper. 



\subsection{6D three-mass-easy hexagons}\label{sec:6d}

Our next example is the 6D three-mass-easy hexagon (a degenerate version of it will later appear in the calculation of 4D double-pentagon)~\cite{DelDuca:2011wh}. The 6D three-mass-easy hexagon is defined as the following pure integral in terms of dual coordinates:
\begin{equation}
    \Omega_1^{(\text{6D})}(i,j,k):=\begin{tikzpicture}[baseline={([yshift=-.5ex]current bounding box.center)},scale=0.18]
        \draw[black,thick] (0,0)--(0,4)--(3.46,6)--(6.93,4)--(6.93,0)--(3.46,-2)--cycle;
        \draw[black,thick] (6.93,4)--(8.66,5);
        \draw[black,thick] (0,4)--(-1.73,5);
        \draw[black,thick] (3.46,-2)--(3.46,-4);
        \filldraw[black] (-1.73,5) node[anchor=south east] {{$i$}};
        \filldraw[black] (8.66,5) node[anchor=south west] {{$j$}};
        \filldraw[black] (3.46,-4) node[anchor=north] {{$k$}};
        \draw[black,thick] (2.46,7.73)--(3.46,6)--(4.46,7.73);
        \filldraw[black] (3.46,7.5) circle [radius=2pt];
        \filldraw[black] (3.98,7.41) circle [radius=2pt];
        \filldraw[black] (2.95,7.41) circle [radius=2pt];
        \draw[black,thick] (-2,0)--(0,0)--(-1,-1.73);
        \filldraw[black] (-1.3,-0.75) circle [radius=2pt];
        \filldraw[black] (-1.48,-0.26) circle [radius=2pt];
        \filldraw[black] (-0.96,-1.15) circle [radius=2pt];
        \draw[black,thick] (8.93,0)--(6.93,0)--(7.93,-1.73);
        \filldraw[black] (8.23,-0.75) circle [radius=2pt];
        \filldraw[black] (7.89,-1.15) circle [radius=2pt];
        \filldraw[black] (8.41,-0.26) circle [radius=2pt];
        \filldraw[black] (3.46,2) node[anchor=center] {{6D}};
    \end{tikzpicture}=\int\frac{{\rm d}^6x_0}{\pi^3}\frac{x_{i,j{+}1}^2x_{j,k{+}1}^2x_{k,i{+}1}^2\sqrt\Delta_9}{x_{0,i}^2x_{0,i{+}1}^2x_{0,j}^2x_{0,j{+}1}^2x_{0,k}^2x_{0,k{+}1}^2}.
\end{equation}
where the normalization in terms of square root of Gram determinant $\Delta_9$ \cite{DelDuca:2011wh} is introduced to make the integral pure:
\begin{equation*}
\begin{split}
    \Delta_9&=(1{-}u_1{-}u_2{-}u_3{+}u_4u_1u_2{+}u_5u_2u_3{+}u_6u_1u_3{-}u_1u_2u_3u_4u_5u_6)^2\\
    &-4u_1u_2u_3(1{-}u_4)(1{-}u_5)(1{-}u_6)
\end{split}
\end{equation*}
with $6$ independent cross ratios
\begin{equation*}
\begin{split}
u_1:=\frac{x_{i{+}1,j{+}1}^2x_{i,k}^2}{x_{i,j{+}1}^2x_{i{+1},k}^2},\ \  u_2:=\frac{x_{j{+}1,k{+}1}^2x_{i,j}^2}{x_{i,j{+}1}^2x_{j,k{+}1}^2},\ \ u_3:=\frac{x_{k{+}1,i{+}1}^2x_{j,k}^2}{x_{i{+}1,k}^2x_{j,k{+}1}^2},\ \ \\
u_4:=\frac{x_{i{+}1,j}^2x_{i,j{+}1}^2}{x_{i,j}^2x_{i{+}1,j{+}1}^2},\ \ 
u_5:=\frac{x_{j{+}1,k}^2x_{j,k{+}1}^2}{x_{j,k}^2x_{j{+}1,k{+}1}^2},\ \ 
u_6:=\frac{x_{i,k{+}1}^2x_{k,i{+}1}^2}{x_{i,k}^2x_{i{+}1,k{+}1}^2}.\ \ 
\end{split}
\end{equation*}

Exactly proceeding as before, we introduce partial Feynman parametrization $y_i:=\xi_ix_{i}+(1-\xi_i)x_{i{+}1}$ on light-like edge $i$ so that $1/(x_{0i}^2x_{0i{+}1}^2)=\int_0^1{\rm d}\xi_i/(x_0-y_i)^4$ (and similarly for $j$ and $k$), it becomes
\begin{equation}
\Omega_1^{(\text{6D})}(i,j,k)=\int^\infty_0{\rm d}^3\xi\int\frac{{\rm d}^6x_0}{\pi^3}\frac{x_{i,j{+}1}^2x_{j,k{+}1}^2x_{k,i{+}1}^2\sqrt{\Delta_9}}{(x_0{-}y_i)^4(x_0{-}y_j)^4(x_0{-}y_k)^4}.
\end{equation}
Similar identity as \eqref{startri} in $6D$ can be deduced by Feynman parametrization:
\begin{equation}
   \int \frac{{\rm d}^6x_0}{\pi^3}\frac1{(x_0{-}y_i)^4(x_0{-}y_j)^4(x_0{-}y_k)^4}=\frac1{(y_i{-}y_j)^2(y_j{-}y_k)^2(y_k{-}y_i)^2},
\end{equation}
which helps us directly perform loop integral to get a rational result! Here we switch to special kinematics where all external momenta (and hence all dual spacetime points $x_a$) lie in 4D subspace of the 6D momentum space, {\it i.e.} only the loop momentum really lives in 6D. This allows us to change to momentum twistor variables in terms of which the square root evaluates nicely to $\sqrt{\Delta_9}=\langle(ijk)\bar i\cap\bar j\cap\bar k\rangle=\langle i\bar j\rangle\langle j\bar k\rangle\langle k\bar i\rangle+\langle i\bar k\rangle\langle k\bar j\rangle\langle j\bar i\rangle$, and we obtain a beautiful formula: 
\begin{equation}\label{6dhexagon1}
    \Omega_1^{(\text{6D})}(i,j,k) = \displaystyle\int_0^\infty{\rm d}^3\tau\frac{\langle(ijk)\bar i\cap\bar j\cap\bar k\rangle}{\langle iXjY\rangle\langle jYkZ\rangle\langle kZiX\rangle},
\end{equation}
with $\{X:=Z_{i-1}-\tau_xZ_{i+1},\ Y:=Z_{j-1}-\tau_yZ_{j+1},\ Z:=Z_{k-1}-\tau_zZ_{k+1}\}$ so that $y_i=(iX)$ and similarly for $y_j,\ y_k$. In this form, we see that the numerator ($\sqrt{\Delta_9}$) makes it possible to rewrite it as ${\rm d}\log$ form:
\begin{equation}\label{6dhexagon}
\boxed{
    \begin{split}
        \Omega_1^{(\text{6D})}(i,j,k) &= \displaystyle\int_0^\infty{\rm d}\tau_z\dfrac{\langle(ijk)\bar i\cap\bar j\cap\bar k\rangle}{\langle kZ\bar i\cap\bar j\rangle\langle kZij\rangle}\begin{tikzpicture}[baseline={([yshift=-.5ex]current bounding box.center)},scale=0.18]
            \draw[black,thick] (0,0)--(5,0)--(6.55,4.76)--(2.50,7.69)--(-1.55,4.76)--cycle;
            \draw[decorate, decoration=snake, segment length=12pt, segment amplitude=2pt, black,thick] (6.55,4.76)--(-1.55,4.76);
            \draw[black,thick] (1.5,9.43)--(2.5,7.69)--(3.5,9.43);
            \filldraw[black] (2.5,9.19) circle [radius=2pt];
            \filldraw[black] (1.99,9.1) circle [radius=2pt];
            \filldraw[black] (3.01,9.1) circle [radius=2pt];
            \draw[black,thick] (-0.21,-1.99)--(0,0)--(-1.83,-0.81);
            \filldraw[black] (-0.88,-1.21) circle [radius=2pt];
            \filldraw[black] (-0.41,-1.44) circle [radius=2pt];
            \filldraw[black] (-1.24,-0.84) circle [radius=2pt];
            \draw[black,thick] (6.83,-0.81)--(5,0)--(5.21,-1.99);
            \filldraw[black] (5.88,-1.21) circle [radius=2pt];
            \filldraw[black] (6.24,-0.84) circle [radius=2pt];
            \filldraw[black] (5.41,-1.44) circle [radius=2pt];
            \draw[black,thick] (6.55,4.76)--(8.45,5.37);
            \draw[black,thick] (-1.55,4.76)--(-3.45,5.37);
            \filldraw[black] (-3.45,5.37) node[anchor=east] {{$i$}};
            \filldraw[black] (8.45,5.37) node[anchor=west] {{$j$}};
            \filldraw[black] (5.21,-1.99) node[anchor=north] {{$k$}};
            \filldraw[black] (-0.21,-1.99) node[anchor=north] {{$Z$}};
        \end{tikzpicture} \\
        &= \displaystyle\int_{\R^3_{\geq 0}}{\rm d}\log\dfrac{\langle kZij\rangle}{\langle kZ\bar i\cap\bar j\rangle}\left({\rm d}\log\dfrac{\langle jYkZ\rangle}{\langle jYi(kZ)\cap\bar i\rangle}{\rm d}\log\dfrac{\langle iXjY\rangle}{\langle iXkZ\rangle}\right).
    \end{split}}
\end{equation}
where the order of integrations is $\tau_X, \tau_Y, \tau_Z$ and after first two we recognized the result is nothing but $\Psi_1$ with $I=(Z k)$! This is a nice identity that expresses the 6D three-mass-easy hexagon as a one-fold ${\rm d}\log$ integral of chiral pentagon. 

We remark that \eqref{6dhexagon1} (rewritten in cross ratios) has been found in \cite{DelDuca:2011wh} using Feynman parametrization. The minor novelty of \eqref{6dhexagon} is in writing it as ${\rm d}\log$ integrals and relating it to the chiral pentagon. We can apply the algorithm to \eqref{6dhexagon}, and indeed we nicely reproduce the symbol of $\Omega_1^{(6D)}(i,j,k)$~\cite{DelDuca:2011wh, Spradlin:2011wp}. In particular we see the formula of \cite{Spradlin:2011wp} directly follow from our algorithm. 

Since we will use the $n=7$ special case (one-mass) of $\Omega_1^{(6D)}$ when studying double-penta ladder, we record its symbol here. The special case we need is $\Omega_{1, {\rm 1-mass}}^{(6D)}:=\Omega_{1, n=7}^{(6D)}(i=1, j=4, k=6)$ where two massive corners degenerate; note $u_5=u_6=0$ and the integral depends on $u_i$ for $i=1,\cdots, 4$. Its symbol reads
\begin{equation}\label{6d1mhex}
\begin{split}
        \mathcal{S}(\Omega_{1, {\rm 1-mass}}^{(6D)})=\mathcal{S}(f_1)\otimes y_1{+}\mathcal{S}(f_2)\otimes y_2{-}\mathcal{S}(f_3)\otimes y_3{-}\mathcal{S}(f_4)\otimes y_4,
    \end{split}
\end{equation}
where we have $4$ dilogarithm functions:
\begin{equation*}
    \begin{split}
        f_1&=\Li_2(1{-}u_3)+\Li_2(1{-}u_4)+\log u_1\log u_2+\log u_3\log u_4,\\
        f_2&=\Li_2(1{-}u_1u_4)+\Li_2(1{-}u_2u_4)-\sum_{i=1}^4\Li_2(1{-}u_i)-\log u_1\log u_2,\\
        f_3&=\Li_2(1{-}u_1u_4)+\Li_2(1{-}u_3)+\log u_1\log u_3+\log u_3\log u_4,\\
        f_4&=\Li_2(1{-}u_2u_4)+\Li_2(1{-}u_3)+\log u_2\log u_3+\log u_3\log u_4,
    \end{split}
\end{equation*}
and the last entries
\begin{equation}\label{ydef}
    \biggl\{y_1=\frac{1{-}x_+}{1{-}x_-},\ \ y_2=\frac{u_4{-}x_+}{u_4{-}x_-},\ \ y_3=\frac{1{-}u_1 x_+}{1{-}u_1 x_-},\ \ y_4=\frac{1{-}u_2 x_+}{1{-}u_2 x_-}\biggr\},
\end{equation}
where we have introduced the combinations involving the square root:
\begin{equation}
x_{\pm}=\frac{-1+u_1+u_2+u_3+u_1u_2u_4\pm\sqrt{\Delta_7}}{2u_1u_2}
\end{equation}
with $\Delta_7=({-}1{+}u_1{+}u_2{+}u_3{+}u_1u_2u_4)^2{-}4u_1u_2u_3(1{-}u_4)$ from the degeneration of $\Delta_9$ as $u_5, u_6 \to 0$.

\section{Penta-ladder integrals}\label{sec.3}

In this section, we move to the simplest generalization of chiral pentagon by attaching a ladder to it. We will see that the Wilson-loop ${\rm d}\log$ form has a recursive structure: the $L$-loop integral is a 2-fold integral of $(L{-}1)$-loop one with shifted kinematics, and growing the ladder is in some sense keep doing certain ``integral transformation". The ${\rm d}\log$ form can again be written in a manifestly DCI form, which makes the computation of the symbol particularly simple.

\subsection{Recursive structure and ${\rm d}\log$ form}

The generic $L$-loop pentaladder integral is defined as

\begin{equation}
        \begin{split}
            \Psi_{L}(i,j;I) &:= \begin{tikzpicture}[baseline={([yshift=-.5ex]current bounding box.center)},scale=0.18]
                \draw[black,thick] (0,0)--(0,5)--(4.76,6.55)--(7.69,2.5)--(4.76,-1.55)--cycle;
                \draw[decorate, decoration=snake, segment length=12pt, segment amplitude=2pt, black,thick] (4.76,6.55)--(4.76,-1.55);
                \draw[black,thick] (9.43,1.5)--(7.69,2.5)--(9.43,3.5);
                \filldraw[black] (9.19,2.5) circle [radius=2pt];
                \filldraw[black] (9.1,1.99) circle [radius=2pt];
                \filldraw[black] (9.1,3.01) circle [radius=2pt];
                \draw[black,thick] (4.76,6.55)--(5.37,8.45);
                \draw[black,thick] (4.76,-1.55)--(5.37,-3.45);
                \draw[black,thick] (0,5)--(-5,5)--(-5,0)--(0,0);
                \draw[thick,densely dashed] (-10,0) -- (-5,0);
                \draw[thick,densely dashed] (-10,5) -- (-5,5);
                \draw[black,thick] (-10,0)--(-10,5)--(-15,5)--(-15,0)--cycle;
                \draw[black,thick] (-16.93,5.52)--(-15,5)--(-15.52,6.93);
                \filldraw[black] (-16.06,6.06) circle [radius=2pt];
                \filldraw[black] (-15.63,6.36) circle [radius=2pt];
                \filldraw[black] (-16.36,5.63) circle [radius=2pt];
                \draw[black,thick] (-16.93,-0.52)--(-15,0)--(-15.52,-1.93);
                \filldraw[black] (-16.06,-1.06) circle[radius=2pt];
                \filldraw[black] (-15.63,-1.36) circle[radius=2pt];
                \filldraw[black] (-16.36,-0.63) circle[radius=2pt];
                \filldraw[black] (5.37,8.45) node[anchor=south west] {{$i$}};
                \filldraw[black] (5.37,-3.45) node[anchor=north west] {{$j$}};
                \filldraw[black] (-15,2.5) node[anchor=east] {{$I$}};
            \end{tikzpicture} \\
            &= \displaystyle\int{\rm d}^{4L}\ell\dfrac{\langle\ell_1\bar i\cap\bar j\rangle\langle Iij\rangle\langle i{-}1ijj{+}1\rangle^{L-1}}{\llangle\ell_1i\rrangle\llangle\ell_1j\rrangle\left[\displaystyle\prod_{a=2}^{L}\langle\ell_{a{-}1}\ell_a\rangle\langle\ell_ai{-}1i\rangle\langle\ell_ajj{+}1\rangle\right]\langle\ell_{L}I\rangle},
        \end{split}
    \end{equation} 
where we use $\{\ell_1,\dots,\ell_L\}$ to denote loop variables from the right-most side to the left-most side, and $I$ is again an arbitrary reference bi-twistor. 

This most general pentaladder integral depends on $8$ external legs just as the chiral pentagon, and the kinematics only depends on the $3$ independent cross ratios, $\{u,v,w\}$, defined in \eqref{crpenta}. We will also denote the $L$-loop pentaladder integral as $\Psi_L(u,v,w)$ and the one-loop chiral pentagon function as $\Psi_1(u,v,w)$, to emphasize their functional dependence on the cross ratios.

To proceed, we can associate these integrals with Feynman diagrams of Wilson loops, but alternatively we can just recycle our result to the pentagon sub-diagram on the right-most side. Parallel to the derivation in section $2$, we introduce $X_1=Z_{i{-}1}{-}\tau_{X_1} Z_{i{+}1}$ and $Y_1=Z_{j{-}1}{-}\tau_{Y_1} Z_{j{+}1}$, use identity \eqref{startri} to integrate the loop $\ell_1$, and rewrite the $L$-loop integral as a two-fold integral of the $(L-1)$-loop one:
 
\begin{equation}\label{pentarecursion}
        \begin{split}
            \Psi_L(i,j,I) &= \displaystyle\int {\rm d}^2\tau\frac{\langle i{-}1ijj{+}1\rangle}{\langle iX_1jY_1\rangle}\int \dfrac{{\rm d}^{4(L-1)}\ell \,\langle\ell_2\bar i\cap\bar j\rangle\langle Iij\rangle\langle i{-}1ijj{+}1\rangle^{L{-}2}}{\llangle\ell_2i\rrangle^{X_1}_{i-1}\llangle\ell_2j\rrangle^{Y_1}_{j{-}1}\left[\displaystyle\prod_{a=3}^{L}\langle\ell_{a{-}1}\ell_a\rangle\langle\ell_ai{-}1i\rangle\langle\ell_ajj{+}1\rangle\right]\langle\ell_{L}I\rangle}\\[-2ex]
&= \displaystyle\int {\rm d} \log\langle i{-}1ijY_1\rangle {\rm d}\log\frac{\langle iX_1jY_1\rangle}{\tau_{X_1}}  \begin{tikzpicture}[baseline={([yshift=-.5ex]current bounding box.center)},scale=0.18]
                \draw[black,thick] (0,0)--(0,5)--(4.76,6.55)--(7.69,2.5)--(4.76,-1.55)--cycle;
                \draw[decorate, decoration=snake, segment length=12pt, segment amplitude=2pt, black,thick] (4.76,6.55)--(4.76,-1.55);
                \draw[black,thick] (9.43,1.5)--(7.69,2.5)--(9.43,3.5);
                \filldraw[black] (9.19,2.5) circle [radius=2pt];
                \filldraw[black] (9.1,1.99) circle [radius=2pt];
                \filldraw[black] (9.1,3.01) circle [radius=2pt];
                \draw[black,thick] (4.76,6.55)--(5.37,8.45);
                \draw[black,thick] (4.76,-1.55)--(5.37,-3.45);
                \draw[thick,densely dashed] (-5,0) -- (0,0);
                \draw[thick,densely dashed] (-5,5) -- (0,5);
                \draw[black,thick] (-5,0)--(-5,5)--(-10,5)--(-10,0)--cycle;
                \draw[black,thick] (-11.93,5.52)--(-10,5)--(-10.52,6.93);
                \filldraw[black] (-11.06,6.06) circle [radius=2pt];
                \filldraw[black] (-10.63,6.36) circle [radius=2pt];
                \filldraw[black] (-11.36,5.63) circle [radius=2pt];
                \draw[black,thick] (-11.93,-0.52)--(-10,0)--(-10.52,-1.93);
                \filldraw[black] (-11.06,-1.06) circle[radius=2pt];
                \filldraw[black] (-10.63,-1.36) circle[radius=2pt];
                \filldraw[black] (-11.36,-0.63) circle[radius=2pt];
                \filldraw[black] (5.37,8.45) node[anchor=south west] {{$i$}};
                \filldraw[black] (5.37,-3.45) node[anchor=north west] {{$j$}};
\filldraw[black] (9.43,3.5) node[anchor=south west] {{$X_1$}};
                \filldraw[black] (9.43,1.5) node[anchor=north west] {{$Y_1$}};
                \filldraw[black] (-10,2.5) node[anchor=east] {{$I$}};
            \end{tikzpicture} \\[-3ex]
&= \displaystyle\int {\rm d} \log\langle i{-}1ijY_1\rangle {\rm d}\log\frac{\langle iX_1jY_1\rangle}{\tau_{X_1}} \tilde\Psi_{L-1}(i,j,I).
        \end{split}
 \end{equation} 
where we introduce the notation $\llangle\ell i\rrangle^{a}_{b}:=\langle\ell i a\rangle\langle\ell i b\rangle$, and the second equality is guaranteed by the identity $\langle\ell_2\bar i\cap\bar j\rangle=\langle\ell_2 (i{-}1\,iX_1)\cap (Y_1j\,j{+}1)\rangle/\tau_{X_1}$; we recognize the $(L{-}1)$-loop integral as a shorter pentaladder but with shifted kinematics: instead of leg $i{+}1$ and $j{-}1$, we have the ``shifted" legs $X_1$, $Y_1$ respectively,   $\tilde{\Psi}_{L-1}=\Psi_{L{-}1}(i{+}1\rightarrow X_1,j{-}1\rightarrow Y_1)$. The upshot is eq.~\eqref{pentarecursion} provides a recursion relation for $\Psi_L$ as $d\log$ integrals of $\tilde{\Psi}_{L{-}1}$. After introducing $X_a=Z_{i-1}-\tau_{X_a}X_{a-1}$ and $Y_{a}=Y_{a-1}-\tau_{Y_a} Z_{j+1}$ for $a=\{2,\dots, L-1\}$, we have
\begin{equation}
\Psi_L(i,j,I)=\int\left[\prod_{a=1}^{L{-}1}{\rm d}\log{\langle i{-}1ijY_a\rangle}{\rm d}\log\frac{\langle iX_ajY_a\rangle}{\tau_{X_a}}\right]\tilde{\Psi}_1.
    \end{equation}
Writing the chiral pentagon itself in its ${\rm d}\log$ representation, we finally arrive at a $2L$-fold ${\rm d}\log$ integral for the $L$-loop pentaladder:
\begin{equation}
\boxed{
\Psi_L(i,j,I)=\int\left[\prod_{a=1}^{L{-}1}{\rm d}\log{\langle i{-}1ijY_a\rangle}{\rm d}\log\frac{\langle iX_ajY_a\rangle}{\tau_{X_a}}\right]{\rm d}\log\frac{\langle jY_{L}I\rangle}{\langle jY_{L}iI\cap\bar{i}\rangle}{\rm d}\log\frac{\langle iX_{L}jY_{L}\rangle}{\langle iX_LI\rangle}.}
\end{equation}
Note that the $(L-1)$ ${\rm d}\log$ 2-forms take a rather different form than the last 2-form for the chiral pentagon.  

Before proceeding, we note that it is trivial to perform half of the integrations, namely all $\tau_{X_a}$-integrals, which fits into the very definition of Goncharov polylogarithms. Here, it is more convenient to use a different parameterization $X_a=s_a Z_{i-1}+(1-s_a)X_{a-1}$ for $0\leq s_a\leq 1$ with $X_{0}:=C Z_{i+1}$, where $C$ is a constant to make sure that all $X_a$ share the same twistor weight. Define $\{t_a:=\langle iX_a I\rangle-\langle i\,i{-}1\,I\rangle\}_{a=1,\dots,L}$ such that each $t_a$ goes form $0$ to $t_{a-1}$, then the $t_a$-integral fits into the definition of Chen's iterated integral \cite{Chen:1977oja}
\[
\int_{t_1=0}^{t_1=C\langle i\,i{+}1\,I\rangle-\langle i\,i{-}1\, I\rangle}\hspace{-3ex}\cdots\int_{t_a=0}^{t_a=t_{a-1}}\hspace{-3ex}\cdots\int_{t_L=0}^{t_L=t_{L-1}}\left[\prod_{a=1}^{L{-}1}{\rm d}\log\frac{\langle iX_ajY_a\rangle}{\langle i{-}1\,I\, X_a\rangle}\right]{\rm d}\log\frac{\langle iX_{L}jY_{L}\rangle}{\langle iX_LI\rangle},
\]
where we see that for any bitwistor $J$ that is independent of $\{X_a\}$,
\[
{\rm d}\log(\langle i X_a J\rangle)={\rm d}\log\left(t_a+
\langle i\,i{-}1\, J\rangle \frac{C\langle i\,i{+}1\, I\rangle-\langle i\,i{-}1\, I\rangle}{C\langle i\,i{+}1\, J\rangle-\langle i\,i{-}1\, J\rangle}
\right),
\]
thus it is completely trivial to integrate to a linear combination of Goncharov polylogarithms. The result will not depend on the choice of $C$ because it is DCI.

\subsection{Recursion relations in a DCI form }\label{sec.3.2}

We see that the order of integrations in \eqref{pentarecursion} is for $X_a$ first and then $Y_a$ for $a=1, \cdots, L$, and since at each step everything is linear,  the computation of the symbol totally straightforward.  Apply the algorithm of section \ref{sec.2.2}  to the recursion
\begin{equation}
\Psi_L(u,v,w)=\int {\rm d} \log\langle i{-}1ijY\rangle\, {\rm d}\log\frac{\langle iXjY\rangle}{\tau_{X}}\Psi_{L-1}(\tilde u,\tilde v,\tilde w)
\end{equation}
with deformed $\tilde u=u(Z_{i{+}1} \rightarrow X, Z_{j{-}1} \rightarrow Y)$, $\tilde{v}$, $\tilde{w}$, we obtain the symbol for $\Psi_L$. The result may not be manifestly DCI, but that can be easily fixed like what we did for $\Psi_1$ in \eqref{I5uvw}. After rescaling $\tau_X$ and $\tau_Y$, we can make the entries of ${\rm d}\log$ forms manifestly DCI: choosing $C_{X}=\frac{\langle i{-}1\,i\,jY\rangle}{\langle i\,i{+}1\,jY\rangle}$ and $C_{Y}=\frac{\langle i{-}1\,i\,j{-}1\,j\rangle}{\langle i{-}1\,i\,j\,j{+}1\rangle}$, we rewrite the recursion as
\begin{equation}\label{recursionpenta}
\boxed{\Psi_{L{+}1}(u,v,w)=\int{\rm d}\log(\tau_{Y}{+}1) \,{\rm d} \log\frac{\tau_{X}{+}1}{\tau_{X}}\Psi_{L}(\tilde u,\tilde v,\tilde w),}
\end{equation}
with deformed cross ratios defined as (for $L\geq 1$):
\begin{equation}
\tilde u=\dfrac{\tau_X\dfrac{\tau_Y+1}{\tau_Y+w}+1}{\dfrac{\tau_X}u\dfrac{\tau_Y+1}{\tau_Y+w}+1},\ \ \tilde v=\dfrac{v(\tau_Y+1)}{v\tau_Y+1},\ \ \tilde w=\dfrac{\tau_X+1}{\tau_X\dfrac{\tau_Y+1}{\tau_Y+w}+1}.
\end{equation}
The recursion can be decomposed into two steps, each increasing weight by $1$:
\begin{equation}\label{penta1}
\Psi_{L{+}\frac12}(u,v,w)=\int {\rm d} \log\frac{\tau_X+1}{\tau_X}\ \Psi_{L}\left(\frac{u(\tau_X+w)}{\tau_X+uw},v,\frac{w(\tau_X+1)}{\tau_X+w}\right)
\end{equation}
and
\begin{equation}\label{penta2}
\Psi_{L{+}1}(u,v,w)=\int{\rm d} \log(\tau_Y+1)\ \Psi_{L{+}\frac12}\left(u,\frac{v(\tau_Y+1)}{ v\tau_Y+1},\frac{\tau_Y+w}{\tau_Y+1}\right)
\end{equation}
where we have defined an ``$(L+\frac12)$-loop'' integral $\Psi_{L+\frac12}(u,v,w)$, whose weight is $2L+1$. We see that from $\Psi_L$ to $\Psi_{L{+}\frac 12}$ and then to $\Psi_{L{+}1}$, we literally apply the algorithm at each step, and obtain the symbol trivially. The functions can also be computed directly, but we emphasize that we do not need to do any integration for the symbol.  

It would be interesting to find a Feynman-diagram interpretation of the odd-weight function $\Psi_{L{+}\frac12}(u,v,w)$. The first example in this family of integrals is $\Psi_{\frac32}(u,v,w)$, whose ${\rm d}\log$ form reads:
\begin{equation}
\begin{split}
\Psi_{\frac32}&=\int {\rm d}\log \frac{\langle iX_1j{-}1j\rangle}{\tau_{X_1}}{\rm d}\log\frac{\langle jY_2I\rangle}{\langle jY_2iI\cap(i{-}1iX_1)\rangle}{\rm d}\log\frac{\langle iX_2jY_2\rangle}{\langle iX_2I\rangle}\\
&= \displaystyle\int{\rm d}\log \frac{\langle iX_1j{-}1j\rangle}{\tau_{X_1}}\times\begin{tikzpicture}[baseline={([yshift=-.5ex]current bounding box.center)},scale=0.18]
            \draw[black,thick] (0,0)--(5,0)--(6.55,4.76)--(2.50,7.69)--(-1.55,4.76)--cycle;
            \draw[decorate, decoration=snake, segment length=12pt, segment amplitude=2pt, black,thick] (6.55,4.76)--(-1.55,4.76);
            \draw[black,thick] (1.5,9.43)--(2.5,7.69)--(3.5,9.43);
            \filldraw[black] (2.5,9.19) circle [radius=2pt];
            \filldraw[black] (1.99,9.1) circle [radius=2pt];
            \filldraw[black] (3.01,9.1) circle [radius=2pt];
            \draw[black,thick] (-0.21,-1.99)--(0,0)--(-1.83,-0.81);
            \filldraw[black] (-0.88,-1.21) circle [radius=2pt];
            \filldraw[black] (-0.41,-1.44) circle [radius=2pt];
            \filldraw[black] (-1.24,-0.84) circle [radius=2pt];
            \draw[black,thick] (6.83,-0.81)--(5,0)--(5.21,-1.99);
            \filldraw[black] (5.88,-1.21) circle [radius=2pt];
            \filldraw[black] (6.24,-0.84) circle [radius=2pt];
            \filldraw[black] (5.41,-1.44) circle [radius=2pt];
            \draw[black,thick] (6.55,4.76)--(8.45,5.37);
            \draw[black,thick] (-1.55,4.76)--(-3.45,5.37);
            \filldraw[black] (-3.45,5.37) node[anchor=east] {{$i$}};
            \filldraw[black] (8.45,5.37) node[anchor=west] {{$j$}};
            \filldraw[black] (1.5,9.43) node[anchor=south east] {{$X_1$}};
            \filldraw[black] (3.5,9.43) node[anchor=south west] {{$j{-}1$}};
            \filldraw[black] (2.5,0) node[anchor=north] {{$I$}};
        \end{tikzpicture}
\end{split}
\end{equation}
with $\{X_1=Z_{i{-}1}-\tau_{X_1}Z_{i{+}1},\ X_2=Z_{i{-}1}-\tau_{X_2}X_1,\ Y_2=Z_{j{-}1}-\tau_{Y_2}Z_{j{+}1}\}$. Note that this is also a one-fold integral of chiral pentagon, similar to the $6D$ three-mass hexagon \eqref{6dhexagon}, but this weight-$3$ integral is quite different. In contrast to that case where we replace bitwistor $I$ to $kZ$, or dual point $x_I\rightarrow z=(kZ)$, here we replace dual point $x_{i{+}1}$ to $x=(i X_1)$ instead, with a different  ${\rm d}\log$ in front. It would be interesting to identity a (partially) $6D$ Feynman integral whose ${\rm d}\log$ representation is $\Psi_{L{+}\frac12}(u,v,w)$.

We remark that \eqref{penta1} and \eqref{penta2} are only one possible choice which makes the ${\rm d} \log$ form explicitly DCI. Other choices of constants $C_{X}$ and $C_{Y}$ give totally different representations. For instance, for the chiral pentagon case, a choice $C_X=\frac{\langle i{-1}\,i\,j{-}1\,j\rangle}{\langle i\,i{+}1\,j{-}1\,j\rangle}$ and $C_Y=\frac{\langle i{-}1\,i\,j{-}1\,j\rangle}{\langle i{-}1\,i\,j\,j{+}1\rangle}$ leads to a representation:
 \begin{equation}
 \Psi_1=\int{\rm d}\log\frac{v \tau_Y+1}{(1-u)\tau_Y+(1-uw)}{\rm d}\log\frac{(1+\tau_Y/w)\tau_X+(1+\tau_Y/(uw))}{\tau_X/(uw)+1}
 \end{equation}
which is distinct from eq.\eqref{I5uvw} obtained in section \ref{sec.2.2}. For the remaining $(L{-}1)-$ loops of penta-ladder, another choice $C_{X}=\frac{\langle i{-}1\,i\,j{-}1\,j\rangle}{\langle i\,i{+}1\,j{-}1\,j\rangle}$ and $C_{Y}=\frac{\langle i{-}1\,i\,j{-}1\,j\rangle}{\langle i{-}1\,i\,j\,j{+}1 \rangle}$ gives a different recursion for the pentaladder:
\begin{equation}\label{pentaladderRecNew}
\Psi_{L{+}1}(u,v,w)=\int{\rm d}\log(\tau_Y{+}1)\,{\rm d}\log\frac{\tau_X(1{+}\tau_Y/w){+}(1{+}\tau_Y)}{\tau_X}\Psi_{L}(\tilde{u},\tilde{v},\tilde{w})
\end{equation}
where the deformed cross ratios now read
\begin{equation}\label{pentaladderRecNewRatio}
\tilde{u}=\frac{1+\tau_X/w}{1+\tau_X/(uw)},\ \ \tilde{v}=\frac{v(1+\tau_Y)}{1+\tau_Y v},\ \ \tilde{w}=\frac{(1{+}\tau_Y){+}\tau_X(1{+}\tau_Y/w)}{(1+\tau_Y)(1+\tau_X/w)}.
\end{equation}
Recursion \eqref{pentaladderRecNew} has simpler deformation of the cross ratios \eqref{pentaladderRecNewRatio} but the ${\rm d}\log$ forms are a bit  more complicated. It turns out to be useful when we look for relations between WL representation and differential equations for penta-ladders in section \ref{sec.5}. Our original recursion,  \eqref{recursionpenta}, with simpler ${\rm d}\log$ but slightly more complicated deformations, makes the study of symbol structure easier, as we see next. 

\subsection{The symbol of pentaladder integrals}\label{sec.3.3}
In this subsection, we record explicit results for the symbol of penta-ladder at $L=2$ as well as certain structure to any $L$. The recursion relation for $L=2$ reads:
\begin{equation}
\Psi_2(u,v,w)=\int {\rm d} \log(\tau_{Y}+1)\, {\rm d} \log\frac{\tau_{X}+1}{\tau_{X}}\Psi_1(\tilde u,\tilde v,\tilde w).
\end{equation}
A direct computation from the symbol of chiral pentagon \eqref{I5symbol}, gives a compact result
\begin{equation}\label{O2symbol}
\mathcal{S}(\Psi_2(u,v,w))=\frac12\mathcal{S}(q_{uv})\otimes \frac{u v}{(1{-}u)(1{-}v)}{+}\frac12\mathcal{S}(q_{u/v})\otimes \frac{u (1{-}v)}{v(1{-}u)}{+}\mathcal{S}(q_w)\otimes (1{-}w).
\end{equation}
where $q_{uv}$, $q_{u/v}$ and $q_w$ are three weight-$3$ generalized polylogarithm functions, whose explicit symbol can be found in Appendix \ref{resultsq}. Due to the axial symmetry of pentaladder, the functions $q_{uv}$ and $q_w$ are invariant when exchanging $u$ and $v$, while $q_{u/v}$ becomes $-q_{u/v}$ under the exchange. We observe that $9$ letters appear in the alphabet:
\begin{equation}\label{O2alphabet}
\{u,\ v,\ 1{-}u,\ 1{-}v,\ 1{-}w,\ w,\ 1{-}uw,\ 1{-}vw,\ 1{-}u{-}v{+}uvw\}.
\end{equation}
Only the first five, as shown in the result \eqref{O2symbol}, are last entries of the symbol. Moreover, we find that the first two entries of $\mathcal S(\Psi_2)$ (or equivalently those of the $q$ functions), after some recombination, are always the symbol of ``one-loop" functions $\log a\log b$ and $\Li_2(1-a)$, {\it i.e.} they are  $a\otimes b+b\otimes a$ or $a\otimes(1-a)$, where $a, b \in \{u,v,w,uw,vw\}$ are first-entries determined by the physical discontinuities. This should be a general feature, which was first observed in~\cite{CaronHuot:2011ky}.

Extending our discussion beyond $2$ loops, we can trivially obtain the symbol for $L=3$, but the explicit result is too lengthy to be recorded here. We observe that the alphabet of $\mathcal{S}(\Psi_3)$ is again given by \eqref{O2alphabet}, and the last entries are again $\{u,1{-}u,v,1{-}v,1{-}w\}$. We expect that the alphabet for $\mathcal{S}(\Psi_L)$ should consist of these $9$ letters to all loops. 

Furthermore, we claim that the symbol $\mathcal{S}(\Psi_L)$ can always be organized into the form 
\begin{equation}\label{generalO}
    \boxed{\mathcal{S}(\Psi_L(u,v,w))=\frac12\mathcal{S}(q^{(L)}_{uv})\otimes \frac{u v}{(1{-}u)(1{-}v)}{+}\frac12\mathcal{S}(q^{(L)}_{u/v})\otimes \frac{u (1{-}v)}{v(1{-}u)}{+}\mathcal{S}(q^{(L)}_w)\otimes (1{-}w),}
\end{equation}
 {\it i.e.} the last entries always combine to $\frac{u v}{(1-u)(1-v)}$, $\frac{u(1-v)}{v(1-u)}$ and $(1-w)$, and the weight $2L{-}1$ symbols in front are always integrable. The proof of this claim is just a simple induction; first, the deformations in \eqref{penta1} turn the $3$ last entries  of $\Psi_L$ be
\begin{equation}\label{defor}
    \biggl\{\frac{uv(\tau_X{+}w)}{\tau_X(1{-}u)(1{-}v)},\ \frac{u(1{-}v)(\tau_X{+}w)}{\tau_X(1{-}u)v},\ \frac{\tau_X(1-w)}{\tau_X+w}\biggr\}
\end{equation}
and without a full computation, we can determine the last entries from \eqref{defor}. For instance, the first one, which can be decomposed into $\frac{uv}{(1{-}u)(1{-}v)}$, $(\tau_X{+}w)$ and $\tau_X$, gives $\{\frac{uv}{(1{-}u)(1{-}v)},w,0\}$ due to ${\rm d}\log\tau_X$, and $\{\frac{uv}{(1{-}u)(1{-}v)},1-w,1\}$ due to ${\rm d}\log(\tau_X+1)$, according to the second part of the algorithm. Moreover, ${\rm d}\log(\frac{\tau_X+1}{\tau_X})$ only contributes last entries $1$ or $0$, according to the first part of the algorithm. Terms with last entry $1$ vanishes, while terms with last entry $0$ diverge, but they add up to zero as $\Psi_{L+\frac 1 2}$ itself is finite. As a result, the algorithm shows that function $\Psi_{L+\frac 1 2 }(u,v,w)$ have $4$ last entries in its symbol, after recombination:
\begin{equation}\label{O52last}
\biggl\{\frac{uv}{(1{-}u)(1{-}v)},\ \frac{u(1-v)}{v(1-u)},\ w,\ 1-w\biggr\}.
\end{equation}
Now it is exactly the same computation to determine the last entries of $\mathcal{S}(\Psi_{L+1})$ from \eqref{penta2} and \eqref{O52last}, {\it i.e.} we firstly deformed the $4$ last entries as
\begin{equation}
    \biggl\{\frac{uv(\tau_Y{+}1)}{(1{-}u)(1{-}v)},\ \frac{u(1{-}v)}{v(1{-}u)(\tau_Y{+}1)},\ \frac{\tau_Y+w}{\tau_Y+1},\ \frac{1-w}{\tau_Y+1}\biggr\}
\end{equation}
then following the algorithm to get the results. The upshot is that last entries of $\mathcal{S}(\Psi_{L+1})$ are exactly given by $\frac{u v}{(1-u)(1-v)}$, $\frac{u(1-v)}{v(1-u)}$ and $(1-w)$. 
This gives a inductive proof that to all loops, last entries of $\mathcal{S}(\Psi_L(u,v,w))$ remain invariant for $L>1$. 

Moreover, we have also shown that weight $2(L{+}1)$ symbols in front of the three last entries are integrable. According to our algorithm, after last entries are computed, the next step is to integrate  $\{\mathcal{S}(q^{(L)}_{uv}),\mathcal{S}(q^{(L)}_{u/v}),\mathcal{S}(q^{(L)}_{w})\}$ in front recursively.  Recombinations of the integrated results then turn out to be $\{\mathcal{S}(q^{(L{+}1)}_{uv}),\mathcal{S}(q^{(L{+}1)}_{u/v}),\mathcal{S}(q^{(L{+}1)}_{w})\}$ at $L{+}1$ loops, which are of course integrable. So we finally arrive at the result \eqref{generalO}. 

One more byproduct from the induction is that we also determined last entries of the odd-weight ones $\mathcal{S}(\Psi_{L{+}\frac12})$ for $L\geq2$, which are always given by \eqref{O52last}. Note that, since recursion \eqref{penta1} breaks the axial symmetry, odd-weight functions do not remain invariant when exchanging $u$ and $v$. Once the second step \eqref{penta2} is performed, the symmetry recovers.

\section{Generalized penta-ladder integrals}\label{sec.4}

Given the success for penta-ladder integrals, it is natural to wonder if one can obtain WL ${\rm d} \log$ representation for more general integrals. One could either search for more examples which correspond to Feynman diagrams of (super-)WL, or keep applying partial Feynman parametrization to other types of integrals. We do not know how to proceed in the most systematic way in either direction, but it is clear that integrals admitting WL representation are ubiquitous. We can consider any $L$-loop integral with $L'$ ``terminal" chiral-pentagon sub-diagrams, {\it i.e.} their three corners only have external legs; then by rewriting each of them as a two-fold $d\log$ integral, one arrives at $2L'$-fold ${\rm d} \log$ integral acting on some $L-L'$ diagram. If there are new ``terminal" chiral pentagons, we can proceed to reduce the diagram to lower loops, and so on. This is a general strategy of relating higher-loop integrals as ${\rm d} \log$ integrals of lower-loop ones, and clearly it works for any one-loop sub-diagrams that can be written as ${\rm d} \log$ forms, {\it e.g.} four-mass box integral (or 6D three-mass-easy hexagon). 

For example, for any two-loop pure integral with a chiral pentagon with loop momentum $\ell'$, we can immediately write it as two-fold ${\rm d} \log$ integral of a one-loop integral, where in the remaining loop, the middle propagator $1/\langle \ell' \ell\rangle$ is replaced with $1/\langle \ell i X\rangle\langle \ell j Y\rangle$ corresponding to choosing $I=\ell$ for the pentagon integration. 
In this way, the double-pentagon becomes two-fold ${\rm d} \log$ integral of a hexagon, and the penta-box becomes two-fold ${\rm d} \log$ integral of a pentagon. Similarly, we can write the three-loop integral for MHV amplitudes with chiral pentagons on both sides~\cite{ArkaniHamed:2010gh}, as four-fold ${\rm d}\log$ integral of an octagon; even more generally, any $L$-loop integrals with $(L{-}1)$ chiral pentagon ``handles" attached to the middle loop, can be written as $2(L{-}1)$-fold integrals of some one-loop integral. The remaining one-loop integral can always be evaluated ({\it e.g.} by box expansion) to weight-2 functions, thus in all these cases we arrive at $2(L{-}1)$-fold integrals of some dilogarithms.

To be more concrete, in this section we focus on a simple class of integrals to demonstrate the recursive nature of our strategy. These are what we call generalized penta-ladder integrals, where on one end is the chiral pentagon and on the other end a generic loop. We can recursively use the identity \eqref{startri} to shorten it down to one-loop:
\begin{equation}
        \begin{split}
            &\begin{tikzpicture}[baseline={([yshift=-.5ex]current bounding box.center)},scale=0.18]
                \draw[black,thick] (0,0)--(0,5)--(4.76,6.55)--(7.69,2.5)--(4.76,-1.55)--cycle;
                \draw[decorate, decoration=snake, segment length=12pt, segment amplitude=2pt, black,thick] (4.76,6.55)--(4.76,-1.55);
                \draw[black,thick] (9.43,1.5)--(7.69,2.5)--(9.43,3.5);
                \filldraw[black] (9.19,2.5) circle [radius=2pt];
                \filldraw[black] (9.1,1.99) circle [radius=2pt];
                \filldraw[black] (9.1,3.01) circle [radius=2pt];
                \draw[black,thick] (4.76,6.55)--(5.37,8.45);
                \draw[black,thick] (4.76,-1.55)--(5.37,-3.45);
                \draw[black,thick] (0,5)--(-5,5)--(-5,0)--(0,0);
                \draw[black,thick,densely dashed] (-5,5)--(-10,5);
                \draw[black,thick,densely dashed] (-5,0)--(-10,0);
                \draw[black,thick] (-15,5)--(-10,5)--(-10,0)--(-15,0);
                \draw[black,thick] (-20,0) arc (-135:135:3.54);
                \draw[black,thick] (-20.92,3.42) arc (165:195:3.54);
                \draw[black,thick] (-17.5,6.04)--(-17.5,8.04);
                \draw[black,thick] (-17.5,-1.04)--(-17.5,-3.04);
                \filldraw[black] (-19.43,7.15) circle [radius=2pt];
                \filldraw[black] (-20.21,6.75) circle [radius=2pt];
                \filldraw[black] (-18.59,7.42) circle [radius=2pt];
                \filldraw[black] (-19.43,-2.15) circle [radius=2pt];
                \filldraw[black] (-20.21,-1.75) circle [radius=2pt];
                \filldraw[black] (-18.59,-2.42) circle [radius=2pt];
                \filldraw[black] (5.37,8.45) node[anchor=south west] {{$i$}};
                \filldraw[black] (5.37,-3.45) node[anchor=north west] {{$j$}};
                \filldraw[black] (-17.5,8.04) node[anchor=south] {{$i{-}1$}};
                \filldraw[black] (-17.5,-3.04) node[anchor=north] {{$j{+}1$}};
            \end{tikzpicture} \\[-4ex]
            &= \displaystyle\int_{\R^2_{\geq 0}}{\rm d}\log\langle i{-}1ijY_1\rangle{\rm d}\log\frac{\langle iX_1jY_1\rangle}{\tau_{X_0}}\times\begin{tikzpicture}[baseline={([yshift=-.5ex]current bounding box.center)},scale=0.18]
                \draw[black,thick] (-5,0)--(-5,5)--(-0.24,6.55)--(2.69,2.5)--(-0.24,-1.55)--cycle;
                \draw[decorate, decoration=snake, segment length=12pt, segment amplitude=2pt, black,thick] (-0.24,6.55)--(-0.24,-1.55);
                \draw[black,thick] (4.43,1.5)--(2.69,2.5)--(4.43,3.5);
                \filldraw[black] (4.19,2.5) circle [radius=2pt];
                \filldraw[black] (4.1,1.99) circle [radius=2pt];
                \filldraw[black] (4.1,3.01) circle [radius=2pt];
                \draw[black,thick] (-0.24,6.55)--(0.37,8.45);
                \draw[black,thick] (-0.24,-1.55)--(0.37,-3.45);
                \draw[black,thick,densely dashed] (-5,5)--(-10,5);
                \draw[black,thick,densely dashed] (-5,0)--(-10,0);
                \draw[black,thick] (-15,5)--(-10,5)--(-10,0)--(-15,0);
                \draw[black,thick] (-20,0) arc (-135:135:3.54);
                \draw[black,thick] (-20.92,3.42) arc (165:195:3.54);
                \draw[black,thick] (-17.5,6.04)--(-17.5,8.04);
                \draw[black,thick] (-17.5,-1.04)--(-17.5,-3.04);
                \filldraw[black] (-19.43,7.15) circle [radius=2pt];
                \filldraw[black] (-20.21,6.75) circle [radius=2pt];
                \filldraw[black] (-18.59,7.42) circle [radius=2pt];
                \filldraw[black] (-19.43,-2.15) circle [radius=2pt];
                \filldraw[black] (-20.21,-1.75) circle [radius=2pt];
                \filldraw[black] (-18.59,-2.42) circle [radius=2pt];
                \filldraw[black] (0.37,8.45) node[anchor=south west] {{$i$}};
                \filldraw[black] (0.37,-3.45) node[anchor=north west] {{$j$}};
                \filldraw[black] (-17.5,8.04) node[anchor=south] {{$i{-}1$}};
                \filldraw[black] (-17.5,-3.04) node[anchor=north] {{$j{+}1$}};
                \filldraw[black] (4.43,3.5) node[anchor=west] {{$X_1$}};
                \filldraw[black] (4.43,1.5) node[anchor=west] {{$Y_1$}};
            \end{tikzpicture} \\[-3ex]
            &= \displaystyle\int_{\R^{2(L{-}1)}_{\geq 0}}\left[\prod_{a=1}^{L{-}1}{\rm d}\log\langle i{-}1ijY_a\rangle{\rm d}\log\frac{\langle iX_ajY_a\rangle}{\tau_{X_a}}\right]\times\begin{tikzpicture}[baseline={([yshift=-.5ex]current bounding box.center)},scale=0.18]
                \draw[black,thick] (-15,0)--(-12.5,2.5)--(-15,5); 
                \draw[decorate, decoration=snake, segment length=12pt, segment amplitude=2pt, black,thick] (-15,0)--(-15,5);
                \draw[black,thick] (-10.76,1.5)--(-12.5,2.5)--(-10.76,3.5);
                \filldraw[black] (-11,2.5) circle [radius=2pt];
                \filldraw[black] (-11.09,1.99) circle [radius=2pt];
                \filldraw[black] (-11.09,3.01) circle [radius=2pt];
                \draw[black,thick] (-15,5)--(-13.59,6.41);
                \draw[black,thick] (-15,0)--(-13.59,-1.41);
                \draw[black,thick] (-20,0) arc (-135:-45:3.54);
                \draw[black,thick] (-15,5) arc (45:135:3.54);
                \draw[black,thick] (-20.92,3.42) arc (165:195:3.54);
                \draw[black,thick] (-17.5,6.04)--(-17.5,8.04);
                \draw[black,thick] (-17.5,-1.04)--(-17.5,-3.04);
                \filldraw[black] (-19.43,7.15) circle [radius=2pt];
                \filldraw[black] (-20.21,6.75) circle [radius=2pt];
                \filldraw[black] (-18.59,7.42) circle [radius=2pt];
                \filldraw[black] (-19.43,-2.15) circle [radius=2pt];
                \filldraw[black] (-20.21,-1.75) circle [radius=2pt];
                \filldraw[black] (-18.59,-2.42) circle [radius=2pt];
                \filldraw[black] (-13.59,6.41) node[anchor=south west] {{$i$}};
                \filldraw[black] (-13.59,-1.41) node[anchor=north west] {{$j$}};
                \filldraw[black] (-17.5,8.04) node[anchor=south] {{$i{-}1$}};
                \filldraw[black] (-17.5,-3.04) node[anchor=north] {{$j{+}1$}};
                \filldraw[black] (-10.76,3.5) node[anchor=west] {{$X_{L{-}1}$}};
                \filldraw[black] (-10.76,1.5) node[anchor=west] {{$Y_{L{-}1}$}};
            \end{tikzpicture}
        \end{split}
\end{equation}
where proper numerator for the left-most loop is needed to make the integral pure. Eventually, the remaining one-loop integral is a combination of the left-most loop with the chiral pentagon ``handle" that is shifted: $Z_{i+1}\to X_{L{-}1}$ and $Z_{j-1}\to Y_{L{-}1}$.

We leave more systematic study WL representation and evaluation of generalized penta-ladder integrals, as well as other cases to future works. Note that unlike the penta-ladder case, generally the one-loop integral does not admit any two-form representation, and moreover in most cases it contains square roots from {\it e.g.} four-mass boxes, which makes it necessary to use rationalization. Such integrals are generically algebraic functions of momentum twistors. For example, the most general double-penta ladder depends on $8$ external legs
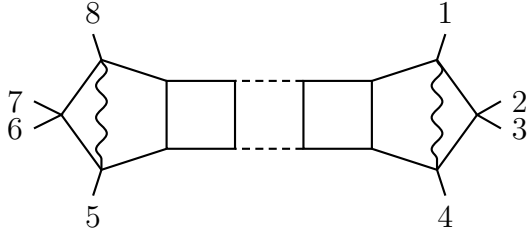
\begin{center} 
\begin{tikzpicture}[baseline={([yshift=-.5ex]current bounding box.center)},scale=0.18]
        \draw[black,thick] (0,0)--(0,5)--(4.76,6.55)--(7.69,2.5)--(4.76,-1.55)--cycle;
        \draw[black,thick] (-15,5)--(-19.76,6.55)--(-22.69,2.5)--(-19.76,-1.55)--(-15,0);
        \draw[decorate, decoration=snake, segment length=12pt, segment amplitude=2pt, black,thick] (4.76,6.55)--(4.76,-1.55);
        \draw[decorate, decoration=snake, segment length=12pt, segment amplitude=2pt, black,thick] (-19.76,6.55)--(-19.76,-1.55);
        \draw[black,thick] (9.43,1.5)--(7.69,2.5)--(9.43,3.5);
        \draw[black,thick] (4.76,6.55)--(5.37,8.45);
        \draw[black,thick] (4.76,-1.55)--(5.37,-3.45);
        \draw[black,thick] (0,5)--(-5,5)--(-5,0)--(0,0);
        \draw[black,thick,densely dashed] (-5,5)--(-10,5);
        \draw[black,thick,densely dashed] (-5,0)--(-10,0);
        \draw[black,thick] (-10,0)--(-10,5)--(-15,5)--(-15,0)--cycle;
        \draw[black,thick] (-19.76,6.55)--(-20.37,8.45);
        \draw[black,thick] (-19.76,-1.55)--(-20.37,-3.45);
        \draw[black,thick] (-24.69,3.5)--(-22.69,2.5)--(-24.69,1.5);
        \filldraw[black] (-20.37,8.45) node[anchor=south] {{8}};
        \filldraw[black] (5.37,8.45) node[anchor=south] {{1}};
        \filldraw[black] (9.43,3.5) node[anchor=west] {{2}};
        \filldraw[black] (9.43,1.5) node[anchor=west] {{3}};
        \filldraw[black] (5.37,-3.45) node[anchor=north] {{4}};
        \filldraw[black] (-20.37,-3.45) node[anchor=north] {{5}};
        \filldraw[black] (-24.69,1.5) node[anchor=east] {{6}};
        \filldraw[black] (-24.69,3.5) node[anchor=east] {{7}};
    \end{tikzpicture}
\end{center}
whose result contains the square root $\sqrt{(1-u-v)^2-4uv}$, where $\{u,v\}$ are the two cross ratios of four mass box $I_4(2,4,6,8)$\footnote{The most generic double-pentagon integral depends on $12$ legs, which contains $16$ such square roots. More details of studying double-pentagon integrals can be found in \cite{He:2021}.}. Since we will focus on evaluation of WL ${\rm d} \log$ forms that are rational, in the rest of the section we restrict ourselves to the simplest double-penta ladder of $n=7$, which does not contain any square root. 
    
\subsection{Double-penta-ladders as ${\rm d}\log$ integrals }\label{sec.4.1}
The seven-point double-penta-ladder is defined as
\begin{equation}\label{doublepentaladder7}
\begin{split}
        \Omega_L(1,4,5,7)&=\begin{tikzpicture}[baseline={([yshift=-.5ex]current bounding box.center)},scale=0.18]
        \draw[black,thick] (0,0)--(0,5)--(4.76,6.55)--(7.69,2.5)--(4.76,-1.55)--cycle;
        \draw[black,thick] (-15,5)--(-19.76,6.55)--(-22.69,2.5)--(-19.76,-1.55)--(-15,0);
        \draw[decorate, decoration=snake, segment length=12pt, segment amplitude=2pt, black,thick] (4.76,6.55)--(4.76,-1.55);
        \draw[decorate, decoration=snake, segment length=12pt, segment amplitude=2pt, black,thick] (-19.76,6.55)--(-19.76,-1.55);
        \draw[black,thick] (9.43,1.5)--(7.69,2.5)--(9.43,3.5);
        \draw[black,thick] (4.76,6.55)--(5.37,8.45);
        \draw[black,thick] (4.76,-1.55)--(5.37,-3.45);
        \draw[black,thick] (0,5)--(-5,5)--(-5,0)--(0,0);
        \draw[black,thick,densely dashed] (-5,5)--(-10,5);
        \draw[black,thick,densely dashed] (-5,0)--(-10,0);
        \draw[black,thick] (-10,0)--(-10,5)--(-15,5)--(-15,0)--cycle;
        \draw[black,thick] (-19.76,6.55)--(-20.37,8.45);
        \draw[black,thick] (-19.76,-1.55)--(-20.37,-3.45);
        \draw[black,thick] (-22.69,2.5)--(-24.69,2.5);
        \filldraw[black] (-20.37,8.45) node[anchor=south] {{7}};
        \filldraw[black] (5.37,8.45) node[anchor=south] {{1}};
        \filldraw[black] (9.43,3.5) node[anchor=west] {{2}};
        \filldraw[black] (9.43,1.5) node[anchor=west] {{3}};
        \filldraw[black] (5.37,-3.45) node[anchor=north] {{4}};
        \filldraw[black] (-20.37,-3.45) node[anchor=north] {{5}};
        \filldraw[black] (-24.69,2.5) node[anchor=east] {{6}};
    \end{tikzpicture}\\
        &=\displaystyle\int{\rm d}^{4L}\ell\dfrac{\langle\ell_1\bar 1\cap\bar 4\rangle\langle\ell_2\bar 5\cap\bar 7\rangle\langle 1457\rangle^{L-1}}{\llangle\ell_11\rrangle\llangle\ell_14\rrangle\left[\displaystyle\prod_{a=2}^{L}\langle\ell_{a{-}1}\ell_a\rangle\langle\ell_a71\rangle\langle\ell_a45\rangle\right]\llangle\ell_L5\rrangle\llangle\ell_L7\rrangle}
\end{split}
\end{equation}
whose symbol entries are all rational DCI variables. Such integrals depend on $4$ independent cross ratios, which throughout this section we choose to be:
\begin{equation}
u_1=\frac{\langle1245\rangle\langle5671\rangle}{\langle1256\rangle\langle4571\rangle},\ 
u_2=\frac{\langle3471\rangle\langle4567\rangle}{\langle3467\rangle\langle4571\rangle},\ 
u_3=\frac{\langle1267\rangle\langle3456\rangle}{\langle1256\rangle\langle3467\rangle},\  
u_4=\frac{\langle1234\rangle\langle4571\rangle}{\langle1245\rangle\langle3471\rangle}.
\end{equation}

Following the general discussion above, after we introduce integration variables recursively, $X_1=Z_7-\tau_{X_1} Z_2$, $Y_1=Z_3-\tau_{Y_1}Z_5$, and $\{X_a=Z_7-\tau_{X_{a}}X_{a{-}1},Y_a=Y_{a{-}1}-\tau_{Y_{a}}Z_5\}_{a=2,\dots,L-1}$, $\Omega_L(1,4,5,7)$ can be rewritten as a $2(L-1)$-fold integral:
\begin{equation}\label{rcdouble}
\boxed{\Omega_L(1,4,5,7)=\int\prod_{a=1}^{L{-}1}{\rm d}\log{\langle 147Y_a\rangle}{\rm d}\log\frac{\langle1X_a4Y_a\rangle}{\tau_{X_a}}\times\begin{tikzpicture}[baseline={([yshift=-.5ex]current bounding box.center)},scale=0.18]
            \draw[black,thick] (0,0)--(4,0)--(6,3.46)--(4,6.93)--(0,6.93)--(-2,3.46)--cycle;
            \draw[black,thick] (0,6.93)--(-1,8.66);
            \draw[black,thick] (4,6.93)--(5,8.66);
            \draw[black,thick] (7.74,2.46)--(6,3.46)--(7.74,4.46);
            \draw[black,thick] (4,0)--(5,-1.73);
            \draw[black,thick] (0,0)--(-1,-1.73);
            \draw[black,thick] (-2,3.46)--(-4,3.46);
            \draw[decorate, decoration=snake, segment length=12pt, segment amplitude=2pt, black,thick] (0,7)--(0,0);
            \draw[decorate, decoration=snake, segment length=12pt, segment amplitude=2pt, black,thick] (4,0)--(4,6.93);
            \filldraw[black] (-1,8.66) node[anchor=south east] {{7}};
            \filldraw[black] (5,8.66) node[anchor=south west] {{1}};
            \filldraw[black] (7.74,4.46) node[anchor=west] {{$X_{L{-}1}$}};
            \filldraw[black] (7.74,2.46) node[anchor=west] {{$Y_{L{-}1}$}};
            \filldraw[black] (5,-1.73) node[anchor=north west] {{4}};
            \filldraw[black] (0,-1.73) node[anchor=north east] {{5}};
            \filldraw[black] (-4,3.46) node[anchor=east] {{6}};
        \end{tikzpicture}.}
\end{equation}
Because of this, from now on we denote the $7$-point hexagon as $\Omega_1$, which serves as the starting point of our recursion relation for double-penta-ladders. Rewritten in DCI form, \eqref{rcdouble} become almost identical to that for penta-ladder, and it can be naturally decomposed into two steps like \eqref{penta1} and \eqref{penta2}:
\begin{equation}\label{double1}
\begin{split}
\Omega_{L+\frac12}(u_1,u_2,u_3,u_4)&=\int {\rm d}\log\frac{\tau_X{+}1}{\tau_X}\Omega_L\biggl(\frac{u_1(\tau_X{+}u_4)}{\tau_X{+}u_1 u_4},u_2,\frac{\tau_X u_3}{\tau_X{+}u_1 u_4},\frac{u_4(\tau_X{+}1)}{\tau_X{+}u_4}\biggr),\\
\Omega_{L{+}1}(u_1,u_2,u_3,u_4)&=\int {\rm d}\log(\tau_Y{+}1)\Omega_{L+\frac12}\biggl(u_1,\frac{u_2(\tau_Y{+}1)}{u_2\tau_Y{+}1},\frac{u_3}{1{+}\tau_Y u_2},\frac{\tau_Y{+}u_4}{\tau_Y{+}1}\biggr),
\end{split}
\end{equation}
where we have introduced the odd-weight integrals $\Omega_{L{+}\frac 12}$ as well. Since the original hexagon integral is well known
 \begin{equation}
        \begin{split}
     \begin{tikzpicture}[baseline={([yshift=-.5ex]current bounding box.center)},scale=0.18]
                \draw[black,thick] (0,0)--(4,0)--(6,3.46)--(4,6.93)--(0,6.93)--(-2,3.46)--cycle;
                \draw[black,thick] (0,6.93)--(-1,8.66);
                \draw[black,thick] (4,6.93)--(5,8.66);
                \draw[black,thick] (7.74,2.46)--(6,3.46)--(7.74,4.46);
                \draw[black,thick] (4,0)--(5,-1.73);
                \draw[black,thick] (0,0)--(-1,-1.73);
                \draw[black,thick] (-2,3.46)--(-4,3.46);
                \draw[decorate, decoration=snake, segment length=12pt, segment amplitude=2pt, black,thick] (0,7)--(0,0);
                \draw[decorate, decoration=snake, segment length=12pt, segment amplitude=2pt, black,thick] (4,0)--(4,6.93);
                \filldraw[black] (-1,8.66) node[anchor=south east] {{7}};
                \filldraw[black] (5,8.66) node[anchor=south west] {{1}};
                \filldraw[black] (7.74,4.46) node[anchor=west] {{2}};
                \filldraw[black] (7.74,2.46) node[anchor=west] {{3}};
                \filldraw[black] (5,-1.73) node[anchor=north west] {{4}};
                \filldraw[black] (0,-1.73) node[anchor=north east] {{5}};
                \filldraw[black] (-4,3.46) node[anchor=east] {{6}};
            \end{tikzpicture}&= \log u_1\log u_2-\Li_2(1)+\Li_2(1-u_1)+\Li_2(1-u_2) \\[-4ex]
            &+ \Li_2(1-u_4)-\Li_2(1-u_1u_4)-\Li_2(1-u_2u_4)+\Li_2(1-u_3),
        \end{split}
    \end{equation}
now it is a routine to use the algorithm of section \ref{sec.2.2} and obtain the symbol for $\Omega_L$ recursively. Before proceeding, we observe that the deformation rules for variables $\{u_1,u_2,u_4\}$  in \eqref{double1} are identical to those for $\{u,v,w\}$ in penta-ladder case. Notice $\Omega_1(u_1,u_2,u_3,u_4)=\Psi_1(u_1,u_2,u_4)-\Li_2(1)+\Li_2(1-u_3)$, which reduces the computation to that for penta-ladder, plus a new term which reads: 
\begin{equation}
\Omega_L=\Psi_L(u_1,u_2,u_4)+ \int\prod_{a=1}^{L{-}1}{\rm d}\log(\tau_{Y_{a}}{+}1)\,{\rm d}\log\frac{\tau_{X_a}{+}1}{\tau_{X_a}}(\Li_2(1{-}\tilde{u}_3)-\Li_2(1)),
\end{equation}
where $\tilde u_3$ is defined by the composition of the two deformations in  \eqref{double1}.

We should emphasize that there is nothing stopping us to perform the partial Feynman parametrization on the left side of \eqref{doublepentaladder7}, which gives rise to a slightly different relation:
\begin{equation}\label{wrongdouble}
\Omega_L(1,4,5,7)=\int\prod_{a=1}^{L-1}{\rm d}\log{\langle 457Y_a\rangle}\, {\rm d}\log\frac{\langle5X_a7Y_a\rangle}{\tau_{X_a}}\times\begin{tikzpicture}[baseline={([yshift=-.5ex]current bounding box.center)},scale=0.18]
            \draw[black,thick] (0,0)--(4,0)--(6,3.46)--(4,6.93)--(0,6.93)--(-2,3.46)--cycle;
            \draw[black,thick] (0,6.93)--(-1,8.66);
            \draw[black,thick] (4,6.93)--(5,8.66);
            \draw[black,thick] (7.74,2.46)--(6,3.46)--(7.74,4.46);
            \draw[black,thick] (4,0)--(5,-1.73);
            \draw[black,thick] (0,0)--(-1,-1.73);
            \draw[black,thick] (-4,2.46)--(-2,3.46)--(-4,4.46);
            \draw[decorate, decoration=snake, segment length=12pt, segment amplitude=2pt, black,thick] (0,7)--(0,0);
            \draw[decorate, decoration=snake, segment length=12pt, segment amplitude=2pt, black,thick] (4,0)--(4,6.93);
            \filldraw[black] (-1,8.66) node[anchor=south east] {{7}};
            \filldraw[black] (5,8.66) node[anchor=south west] {{1}};
            \filldraw[black] (7.74,4.46) node[anchor=west] {{$2$}};
            \filldraw[black] (7.74,2.46) node[anchor=west] {{$3$}};
            \filldraw[black] (5,-1.73) node[anchor=north west] {{4}};
            \filldraw[black] (0,-1.73) node[anchor=north east] {{5}};
            \filldraw[black] (-4,4.46) node[anchor=east] {{$Y_{L{-}1}$}};
            \filldraw[black] (-4,2.46) node[anchor=east] {{$X_{L{-}1}$}};
        \end{tikzpicture}.
\end{equation}
Here $X_a=Z_4-\tau_{X_{a}}X_{a{-}1}$, $Y_a=Y_{a{-}1}-\tau_{Y_a}Z_1$, $X_1=Z_4-\tau_{X_1}Z_6$, $Y_1=Z_6-\tau_{Y_1}Z_1$ and the $8$-point hexagon reads:
\begin{equation}
\begin{split}
    \begin{tikzpicture}[baseline={([yshift=-.5ex]current bounding box.center)},scale=0.18]
            \draw[black,thick] (0,0)--(4,0)--(6,3.46)--(4,6.93)--(0,6.93)--(-2,3.46)--cycle;
            \draw[black,thick] (0,6.93)--(-1,8.66);
            \draw[black,thick] (4,6.93)--(5,8.66);
            \draw[black,thick] (7.74,2.46)--(6,3.46)--(7.74,4.46);
            \draw[black,thick] (4,0)--(5,-1.73);
            \draw[black,thick] (0,0)--(-1,-1.73);
            \draw[black,thick] (-4,2.46)--(-2,3.46)--(-4,4.46);
            \draw[decorate, decoration=snake, segment length=12pt, segment amplitude=2pt, black,thick] (0,7)--(0,0);
            \draw[decorate, decoration=snake, segment length=12pt, segment amplitude=2pt, black,thick] (4,0)--(4,6.93);
            \filldraw[black] (-1,8.66) node[anchor=south east] {{8}};
            \filldraw[black] (5,8.66) node[anchor=south west] {{1}};
            \filldraw[black] (7.74,4.46) node[anchor=west] {{$2$}};
            \filldraw[black] (7.74,2.46) node[anchor=west] {{$3$}};
            \filldraw[black] (5,-1.73) node[anchor=north west] {{4}};
            \filldraw[black] (0,-1.73) node[anchor=north east] {{5}};
            \filldraw[black] (-4,4.46) node[anchor=east] {{$7$}};
            \filldraw[black] (-4,2.46) node[anchor=east] {{$6$}};
        \end{tikzpicture}&=\log(u_{1,4,5,8})\log(u_{2,5,6,1})\\[-6ex]
        &+\Li_2(1-u_{2,5,6,1})-\Li_2(1-u_{6,1,2,4})+\Li_2(1-u_{2,4,5,1})\\
        &-\Li_2(1-u_{2,5,6,8})+\tilde{I}_4(2,4,6,8)-\Li_2(1-u_{2,4,5,8})\\
        &+\Li_2(1-u_{1,5,6,8})-\Li_2(1-u_{6,8,1,4})+\Li_2(1-u_{1,4,5,8}),
\end{split}
\end{equation}
where \[u_{a,b,c,d}=\frac{\langle a{-}1\ a\ b{-}1\ b\rangle\langle c{-}1\ c\ d{-}1\ d\rangle}{\langle a{-}1\ a\ c{-}1\ c\rangle\langle b{-}1\ b\ d{-}1\ d\rangle}\]
and the modified four mass box
\[\tilde{I}_4(2,4,6,8)=\frac{1{-}u_{2,4,6,8}{-}u_{8,2,4,6}}{\sqrt{(1{-}u_{2,4,6,8}{-}u_{8,2,4,6})^2{-}4u_{2,4,6,8}u_{8,2,4,6}}}I_4(2,4,6,8)-\frac12\log(u_{2,4,6,8})\log(u_{8,2,4,6}).\]
The integrand of \eqref{wrongdouble} therefore contains a non-trivial square root $\Delta_4=\sqrt{(1{-}u{-}v)^2{-}4uv}$ with $u=\frac{\langle5X7Y\rangle\langle1234\rangle}{\langle5X12\rangle\langle7Y34\rangle}$ and $v=\frac{\langle5X34\rangle\langle7Y12\rangle}{\langle5X12\rangle\langle7Y34\rangle}$. However, after rationalizing the square root and performing the $2(L{-}1)$-fold integration, function $\Omega_L$ contains only four independent ``square roots''  
\begin{equation*}
\{\Delta_4(X{\to}4,Y{\to}6),\ \Delta_4(X{\to}4,Y{\to}1),\ \Delta_4(X{\to}6,Y{\to}6),\ \Delta_4(X{\to}6,Y{\to}1)\}.
\end{equation*}
 {\it i.e.} evaluating $\Delta_4$ at the endpoints $x=x_5,x_6$ and $y=x_7,x_1$. In all these four cases $\Delta_4$ degenerates and gives rational results. The complete symbol is rational in terms of momentum twistors, as expected.

\subsection{The symbol of seven-point double-penta ladders}

Let us first present some explicit results for $L=2$ double-penta ladder {\it i.e.} double-pentagon, which can be rewritten in a two-fold integration over hexagon function $\Omega_1$ (we record the definition of the integral again)
\begin{equation}\label{doublepenta}
\begin{split}
        \Omega_2(1,4,5,7)&=\begin{tikzpicture}[baseline={([yshift=-.5ex]current bounding box.center)},scale=0.18]
            \draw[black,thick] (0,0)--(0,5)--(4.76,6.55)--(7.69,2.5)--(4.76,-1.55)--cycle;
            \draw[black,thick] (0,5)--(-4.76,6.55)--(-7.69,2.5)--(-4.76,-1.55)--(0,0);
            \draw[decorate, decoration=snake, segment length=12pt, segment amplitude=2pt, black,thick] (4.76,6.55)--(4.76,-1.55);
            \draw[decorate, decoration=snake, segment length=12pt, segment amplitude=2pt, black,thick] (-4.76,6.55)--(-4.76,-1.55);
            \draw[black,thick] (9.43,1.5)--(7.69,2.5)--(9.43,3.5);
            \draw[black,thick] (4.76,6.55)--(5.37,8.45);
            \draw[black,thick] (4.76,-1.55)--(5.37,-3.45);
            \draw[black,thick] (-4.76,6.55)--(-5.37,8.45);
            \draw[black,thick] (-4.76,-1.55)--(-5.37,-3.45);
            \draw[black,thick] (-7.69,2.5)--(-9.69,2.5);
            \filldraw[black] (-5.37,8.45) node[anchor=south] {{7}};
            \filldraw[black] (5.37,8.45) node[anchor=south] {{1}};
            \filldraw[black] (9.43,3.5) node[anchor=west] {{2}};
            \filldraw[black] (9.43,1.5) node[anchor=west] {{3}};
            \filldraw[black] (5.37,-3.45) node[anchor=north] {{4}};
            \filldraw[black] (-5.37,-3.45) node[anchor=north] {{5}};
            \filldraw[black] (-9.69,2.5) node[anchor=east] {{6}};
        \end{tikzpicture}=\int d^4\ell\frac{\langle\ell_1\bar{1}\cap\bar{4} \rangle\langle\ell_2\bar{5}\cap\bar{7}\rangle\langle1457\rangle}{\llangle\ell_11\rrangle\llangle\ell_14\rrangle\llangle\ell_25\rrangle\llangle\ell_27\rrangle\langle\ell_1\ell_2\rangle}\\[-4.5ex]
        &=\int{\rm d}\log{\langle 147Y\rangle}{\rm d}\log\frac{\langle1X4Y\rangle}{\tau_X}\times\begin{tikzpicture}[baseline={([yshift=-.5ex]current bounding box.center)},scale=0.18]
            \draw[black,thick] (0,0)--(4,0)--(6,3.46)--(4,6.93)--(0,6.93)--(-2,3.46)--cycle;
            \draw[black,thick] (0,6.93)--(-1,8.66);
            \draw[black,thick] (4,6.93)--(5,8.66);
            \draw[black,thick] (7.74,2.46)--(6,3.46)--(7.74,4.46);
            \draw[black,thick] (4,0)--(5,-1.73);
            \draw[black,thick] (0,0)--(-1,-1.73);
            \draw[black,thick] (-2,3.46)--(-4,3.46);
            \draw[decorate, decoration=snake, segment length=12pt, segment amplitude=2pt, black,thick] (0,7)--(0,0);
            \draw[decorate, decoration=snake, segment length=12pt, segment amplitude=2pt, black,thick] (4,0)--(4,6.93);
            \filldraw[black] (-1,8.66) node[anchor=south east] {{7}};
            \filldraw[black] (5,8.66) node[anchor=south west] {{1}};
            \filldraw[black] (7.74,4.46) node[anchor=west] {{$X$}};
            \filldraw[black] (7.74,2.46) node[anchor=west] {{$Y$}};
            \filldraw[black] (5,-1.73) node[anchor=north west] {{4}};
            \filldraw[black] (0,-1.73) node[anchor=north east] {{5}};
            \filldraw[black] (-4,3.46) node[anchor=east] {{6}};
        \end{tikzpicture}.
\end{split}
\end{equation}
The symbol of this integral reads (various $q$ functions are defined in  Appendix \ref{resultsq}):
\begin{equation}\label{Omega2symbol}
\begin{split}
\mathcal{S}(&\Omega_2(u_1,u_2,u_3,u_4))=\mathcal{S}(\Psi_2(u_1,u_2,u_4))+\frac12\mathcal{S}(q_{u_1u_2})\otimes\frac{u_1u_2}{(1{-}u_1)(1{-}u_2)}\\
&+\frac12\mathcal{S}(q_{u_1/u_2})\otimes\frac{u_1(1{-}u_2)}{u_2(1{-}u_1)}
+\frac12\mathcal{S}(q_{u_4})\otimes(1{-}u_4)+\frac12\mathcal{S}(\Omega_{1, {\rm 1-mass}}^{(6D)})\otimes\frac{1-x_+}{1-x_-},
\end{split}
\end{equation}
where $\Omega_{1, {\rm 1-mass}}^{(6D)}$ is the one-mass hexagon in 6d whose symbol, together with definition of $x_\pm$ involving the square root of $\Delta_7$, are given in the end of section~\ref{sec:6d}.

We make some comments on this result. First, the square root in terms of cross ratios, $\sqrt{\Delta_7}$, first appears in the second step of \eqref{double1} for $L=1$. Following the algorithm, after deforming $u_2\rightarrow\tilde u_2=\frac{u_2(\tau_Y+1)}{u_2\tau_Y+1}$ {\it etc.}, we need to reduce all the symbol entries in  products of fractional linear functions for $\tau_Y$.  However, we encountered symbol entries quadratic in $\tau_Y$, $\tau_Y^2+ a \tau_Y+ b$, where $\Delta_7$ is proportional to the discriminant, resulting in this non-trivial square root of cross ratios. Of course, as familiar in dealing with 6D hexagon, this square root becomes rational when using {\it e.g.} momentum twistors variables:
\begin{equation}
\Delta_7=\left(\frac{\langle1267\rangle\langle4567\rangle\langle1345\rangle-\langle1247\rangle\langle1567\rangle\langle3456\rangle}{\langle1256\rangle\langle3467\rangle\langle4571\rangle}\right)^2,
\end{equation} 
thus the result remains rational in terms of momentum twistors. Note $q_{u_1u_2}$, $q_{u_1/u_2}$ and $q_{u_4}$ are weight-$3$ functions whose entries are rational functions of $\{u_1,u_2,u_3,u_4\}$; $q_{u_1u_2}$ and $q_{u_4}$ are symmetric under the exchange of $u_1$ and $u_2$ while $q_{u_1/u_2}$ is antisymmetric. As shown in \eqref{6d1mhex}, $\mathcal{S}(\Omega_{1, {\rm 1-mass}}^{(6D)})$ is the symbol of one mass $6D$ hexagon function \cite{DelDuca:2011jm}, whose last entries depends on $x_\pm$ as well.

Moreover, we find that $\mathcal{S}(\Omega_2)$ has $16$ letters in its alphabet, which are the first $8$ letters of $\Psi_2$, and $8$ new ones as follows
\begin{align*}
\{u_3,\ 1{-}u_3,\ &1{-}u_3{-}u_1u_4,\ 1{-}u_3{-}u_2u_4;\ y_1,\ y_2,\ y_3,\ y_4\}
\end{align*} 
where $y_i$s are defined in \eqref{ydef}. There is a subtlety: note that the ninth letter of $\Psi_2$, $1{-}u_1{-}u_2{+} u_1 u_2 u_4$ appears in ${\cal S}(\Psi_2)$ as well as in the symbol of various $q$ functions, but interestingly they cancel in the final answer!

Finally, following the same logic in \ref{sec.3.3}, we can obtain general structures of $\mathcal{S}(\Omega_L)$ to all loops. We claim that only the following $4$ combinations appear as final entries for double-penta ladders:
\begin{equation}\label{7lasten}
\biggl\{\frac{u_1u_2}{(1{-}u_1)(1{-}u_2)},\ \frac{u_1(1{-}u_2)}{u_2(1{-}u_1)},\ 1{-}u_4,\ \frac{1{-}x_+}{1{-}x_-}\biggr\},
\end{equation}
and we can always put $\mathcal{S}(\Omega_L)$ in the following form:
\begin{equation}\label{generalP}
\boxed{
\begin{split}
\mathcal{S}(&\Omega_L(u_1,u_2,u_3,u_4))=\mathcal{S}(\Psi_L(u_1,u_2,u_4))+\frac12\mathcal{S}(q^{(L)}_{u_1u_2})\otimes\frac{u_1u_2}{(1{-}u_1)(1{-}u_2)}\\
&+\frac12\mathcal{S}(q^{(L)}_{u_1/u_2})\otimes\frac{u_1(1{-}u_2)}{u_2(1{-}u_1)}
+\frac12\mathcal{S}(q^{(L)}_{u_4})\otimes(1{-}u_4)+\frac12\mathcal{S}(q^{(L)}_{\sqrt{\Delta_7}})\otimes\frac{1-x_+}{1-x_-}.
\end{split}}
\end{equation}
where we have weight-$2L{-}1$ functions, $\{q^{(L)}_{u_1/u_2},q^{(L)}_{u_1u_2},q^{(L)}_{u_4}\}$ as well as $q^{(L)}_{\sqrt{\Delta_7}}$ (all of which may have entries that are algebraic functions of cross ratios as well). The proof of this claim is again an induction following our algorithm. As deformations for $\{u_1,u_2,u_4\}$ in \eqref{double1} are exactly the same as what for $\{u,v,w\}$ in \eqref{penta1} and \eqref{penta2}, proof of the first four terms in structure \eqref{generalP} is trivial. Moreover, the entry $\frac{1-x_+}{1-x_-}$ actually stays invariant under each step of the deformation in \eqref{double1}! Therefore, according to the second part of our algorithm, symbol integration over the last term in \eqref{generalP} is just
\begin{equation}\label{Plastterm}
    \frac12\biggl(\int\dif\log{(\tau_Y{+}1)}\,\dif\log\frac{\tau_X{+}1}{\tau_X}\mathcal{S}(\tilde{q}^{(L)}_{\sqrt{\Delta_7}})\biggr)\otimes\frac{1-x_+}{1-x_-}
\end{equation}
where tilde on $q^{(L)}_{\sqrt{\Delta_7}}$ means deforming the cross ratios. So we finish the proof.

In fact, this induction also determines the function $q^{(L)}_{\sqrt{\Delta_7}}$.  According to our notation in \eqref{generalP}, integrated symbol in the bracket of \eqref{Plastterm} is nothing but $\mathcal{S}(q^{(L+1)}_{\sqrt{\Delta_7}})$. \eqref{Plastterm} therefore leads to a recursive definition for functions $q^{(L)}_{\sqrt{\Delta_7}}$
\begin{equation}\label{recursiondelta}
   q^{(L+1)}_{\sqrt{\Delta_7}}(u_1,u_2,u_3,u_4)=\int\dif\log{(\tau_Y{+}1)}\,\dif\log\frac{\tau_X{+}1}{\tau_X}q^{(L)}_{\sqrt{\Delta_7}}(\tilde{u}_1,\tilde{u}_2,\tilde{u}_3,\tilde{u}_4).
\end{equation}

\subsection{${\rm d}\log$ form for the hexagon}

Before ending the section, we remark that our WL representation does not always give $2L$ ${\rm d} \log$'s for $L$-loop integrals. Even for generalized penta-ladders, it is clear that the starting point, which is a one-loop integral, does not always admit a natural two-fold ${\rm d}\log$ integral representation~\footnote{Being weight-$2$ functions, it is always possible to write them as two-fold ${\rm d}\log$ integrals, but in general that would not be natural at all.}. We can already see this in the simplest case beyond the pentagon, namely if we apply partial Feynman parametrization to hexagon integral, we find two-fold integral up to a boundary term, which is a ${\rm d}\log$ integral of a logarithmic function. 

It suffices to look at just the $6$-point hexagon since the $7$-point one works in exactly the same way.  Consider the hexagon
\begin{equation}
        \begin{split}
     \begin{tikzpicture}[baseline={([yshift=-.5ex]current bounding box.center)},scale=0.18]
                \draw[black,thick] (0,0)--(4,0)--(6,3.46)--(4,6.93)--(0,6.93)--(-2,3.46)--cycle;
                \draw[black,thick] (0,6.93)--(-1,8.66);
                \draw[black,thick] (4,6.93)--(5,8.66);
                \draw[black,thick] (6,3.46)--(7.74,3.46);
                \draw[black,thick] (4,0)--(5,-1.73);
                \draw[black,thick] (0,0)--(-1,-1.73);
                \draw[black,thick] (-2,3.46)--(-4,3.46);
                \draw[decorate, decoration=snake, segment length=12pt, segment amplitude=2pt, black,thick] (0,7)--(0,0);
                \draw[decorate, decoration=snake, segment length=12pt, segment amplitude=2pt, black,thick] (4,0)--(4,6.93);
                \filldraw[black] (-1,8.66) node[anchor=south east] {{5}};
                \filldraw[black] (5,8.66) node[anchor=south west] {{6}};
                \filldraw[black] (7.74,3.46) node[anchor=west] {{1}};
                \filldraw[black] (5,-1.73) node[anchor=north west] {{2}};
                \filldraw[black] (0,-1.73) node[anchor=north east] {{3}};
                \filldraw[black] (-4,3.46) node[anchor=east] {{4}};
            \end{tikzpicture}&=\int d^4\ell\frac{\langle\ell\bar{3}\cap\bar{5}\rangle\langle\ell\bar{6}\cap\bar{2}\rangle}{\llangle\ell3\rrangle\llangle\ell5\rrangle\llangle\ell1\rrangle},
        \end{split}
    \end{equation}
and after introducing the parameterization $X=Z_2-\tau_XZ_4$, $Y=Z_4-\tau_YZ_6$, we have
\begin{equation}
    \int{\rm d}^2\tau\int {\rm d}^4\ell\frac{\langle\ell\bar{3}\cap\bar{5}\rangle\langle\ell\bar{6}\cap\bar{2}\rangle}{\langle\ell3X\rangle^2\langle\ell5Y\rangle^2\llangle\ell1\rrangle}.
\end{equation}
Performing the loop integration ${\rm d}^4\ell$ by Feynman parametrization, we are then left with a two-fold integral in $\tau_X$ and $\tau_Y$ as:
\begin{equation}
\begin{split}
\int{\rm d}\log\langle5Y3(61)\cap(234)\rangle\,&{\rm d}\log\frac{\langle3X61\rangle}{\langle3X5Y\rangle}+{\rm d}\log\langle5Y12\rangle\,{\rm d}\log\frac{\langle3X5Y\rangle}{\langle3X1(5Y)\cap(612)\rangle}\\
&-\int{\rm d}^2\tau\frac{\langle1234\rangle\langle1256\rangle\langle2361\rangle\langle4561\rangle}{\langle3X1(5Y)\cap(612)\rangle^2}\log\frac{\langle3X61\rangle\langle5Y12\rangle}{\langle3X12\rangle\langle5Y61\rangle}.
\end{split}
\end{equation}
We see that this is very different from the pentagon case: the integrand is no longer weight-$0$. After integration by part, the weight-$1$ term in this result leads to a boundary term in the ${\rm d}\log$ representation for this integral, so that the final result is still uniformly of weight $2$:
\begin{equation}
\begin{split}
    \begin{tikzpicture}[baseline={([yshift=-.5ex]current bounding box.center)},scale=0.18]
                \draw[black,thick] (0,0)--(4,0)--(6,3.46)--(4,6.93)--(0,6.93)--(-2,3.46)--cycle;
                \draw[black,thick] (0,6.93)--(-1,8.66);
                \draw[black,thick] (4,6.93)--(5,8.66);
                \draw[black,thick] (6,3.46)--(7.74,3.46);
                \draw[black,thick] (4,0)--(5,-1.73);
                \draw[black,thick] (0,0)--(-1,-1.73);
                \draw[black,thick] (-2,3.46)--(-4,3.46);
                \draw[decorate, decoration=snake, segment length=12pt, segment amplitude=2pt, black,thick] (0,7)--(0,0);
                \draw[decorate, decoration=snake, segment length=12pt, segment amplitude=2pt, black,thick] (4,0)--(4,6.93);
                \filldraw[black] (-1,8.66) node[anchor=south east] {{5}};
                \filldraw[black] (5,8.66) node[anchor=south west] {{6}};
                \filldraw[black] (7.74,3.46) node[anchor=west] {{1}};
                \filldraw[black] (5,-1.73) node[anchor=north west] {{2}};
                \filldraw[black] (0,-1.73) node[anchor=north east] {{3}};
                \filldraw[black] (-4,3.46) node[anchor=east] {{4}};
            \end{tikzpicture}&=\int{\rm d}\log\frac{\langle5Y12\rangle}{\langle5Y3(16)\cap(234)\rangle}{\rm d}\log\frac{\langle3X5Y\rangle}{\langle3X61\rangle}\\[-4ex]
            &+\int{\rm d}\log\frac{\langle3X1(5Y)\cap(612)\rangle}{\langle3X61\rangle}\times\log\frac{\langle3X61\rangle\langle5Y12\rangle}{\langle3X12\rangle\langle5Y61\rangle}\biggm|_{Y\to 4}\\
            &=\Li_2(1-u_1)+\Li_2(1-u_2)+\Li_2(1-u_3)-2\Li_2(1)+\log(u_1)\log(u_3)
            \end{split}
\end{equation}
which is the well-known result, with cross ratios
\[u_1=\frac{\langle3456\rangle\langle6123\rangle}{\langle3461\rangle\langle5623\rangle},\quad  u_2=\frac{\langle1234\rangle\langle4561\rangle}{\langle1245\rangle\langle3461\rangle},\quad 
u_3=\frac{\langle2345\rangle\langle5612\rangle}{\langle2356\rangle\langle1245\rangle}. \]

Thus we have the $6$-point double-penta ladder as $2L$ ${\rm d} \log$ integral with a boundary term that is $2L{-}1$ ${\rm d} \log$ integral of a log: 
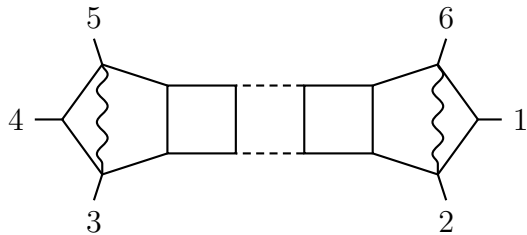
\begin{equation}
    \begin{split}
&\begin{tikzpicture}[baseline={([yshift=-.5ex]current bounding box.center)},scale=0.18]
        \draw[black,thick] (0,0)--(0,5)--(4.76,6.55)--(7.69,2.5)--(4.76,-1.55)--cycle;
        \draw[black,thick] (-15,5)--(-19.76,6.55)--(-22.69,2.5)--(-19.76,-1.55)--(-15,0);
        \draw[decorate, decoration=snake, segment length=12pt, segment amplitude=2pt, black,thick] (4.76,6.55)--(4.76,-1.55);
        \draw[decorate, decoration=snake, segment length=12pt, segment amplitude=2pt, black,thick] (-19.76,6.55)--(-19.76,-1.55);
        \draw[black,thick] (7.69,2.5)--(9.43,2.5);
        \draw[black,thick] (4.76,6.55)--(5.37,8.45);
        \draw[black,thick] (4.76,-1.55)--(5.37,-3.45);
        \draw[black,thick] (0,5)--(-5,5)--(-5,0)--(0,0);
        \draw[black,thick,densely dashed] (-5,5)--(-10,5);
        \draw[black,thick,densely dashed] (-5,0)--(-10,0);
        \draw[black,thick] (-10,0)--(-10,5)--(-15,5)--(-15,0)--cycle;
        \draw[black,thick] (-19.76,6.55)--(-20.37,8.45);
        \draw[black,thick] (-19.76,-1.55)--(-20.37,-3.45);
        \draw[black,thick] (-22.69,2.5)--(-24.69,2.5);
        \filldraw[black] (-20.37,8.45) node[anchor=south] {{5}};
        \filldraw[black] (5.37,8.45) node[anchor=south] {{6}};
        \filldraw[black] (9.43,2.5) node[anchor=west] {{1}};
        \filldraw[black] (5.37,-3.45) node[anchor=north] {{2}};
        \filldraw[black] (-20.37,-3.45) node[anchor=north] {{3}};
        \filldraw[black] (-24.69,2.5) node[anchor=east] {{4}};
    \end{tikzpicture}\\
    &=\int\prod_{a=1}^{L{-}1}{\rm d}\log\langle235Y_a\rangle\,{\rm d}\log\frac{\langle3X_a5Y_a\rangle}{\tau_{X_a}}\times {\rm d}\log\frac{\langle5Y_L12\rangle}{\langle5Y_{L}3(16)\cap(234)\rangle}{\rm d}\log\frac{\langle3X_L5Y_L\rangle}{\langle3X_L61\rangle}\\
    &+\int\prod_{a=1}^{L{-}1}{\rm d}\log\langle235Y_a\rangle\,{\rm d}\log\frac{\langle3X_a5Y_a\rangle}{\tau_{X_a}}{\rm d}\log\frac{\langle3X_L1(5Y_L)\cap(612)\rangle}{\langle3X_L61\rangle}\times\log\biggl(\frac{\langle3X_L61\rangle\langle5Y_L12\rangle}{\langle3X_L12\rangle\langle5Y_L61\rangle}\biggr)\biggm|_{Y_L\to Y_{L{-}1}}
\end{split}
\end{equation}
This result extends not only to $7$-point case, but also to higher-point double-penta ladders where one needs to deal with the square root.

\section{Comments on differential equations and resummation}\label{sec.5}

As we have discussed, our WL ${\rm d}\log$ representation trivializes the evaluation of certain Feynman integrals, and some of these integrals have been studied using other methods, most notably differential equations~\cite{Henn:2013pwa,Henn:2014qga} and even resumming ladders to obtain results at finite coupling~\cite{Caron-Huot:2018dsv}. Here we briefly comment on the way our WL ${\rm d} \log$ representation satisfies differential equations for the ladders, and more importantly their resummation by solving certain integral equations.

\subsection{WL ${\rm d}\log$ representations and differential equations}

It is clear that our method is closely related to solving similar integrals using differential equations~\cite{Henn:2013pwa}, including both second-order equations reducing loop order by one~\cite{Drummond:2010cz}, and first-order ones similar to ${\bar Q}$ equations~\cite{Chicherin:2018rpz,Chicherin:2018ubl}. Whenever we convert a loop to two-fold integral, it must be possible to find a second-order differential operator which annihilates the loop (the difference being the result of this does not involve deformed kinematics). From this point of view, our representation provides the solution to such differential equations without the need of first writing them down and then solving them. 

It would be interesting to study the precise connection of WL ${\rm d}\log$ representation to differential equations in general, but here we content ourselves with checking that our integral formula indeed satisfy the differential equations in~\cite{Drummond:2010cz} for penta-ladder integrals. Given the recursion relation \eqref{pentaladderRecNew} and \eqref{pentaladderRecNewRatio} for penta-ladder, it is easy to derive that they satisfy differential equation found in \cite{Drummond:2010cz}:
\begin{equation}\label{pentaladderDiffEq}
    (1-u-v+uvw)uv\partial_u\partial_v\Psi_{L+1}(u,v,w)=\Psi_L(u,v,w).
\end{equation}
The key observation is that the differential operator $\mathcal D:=(1-u-v+uvw)uv\partial_u\partial_v$ commutes with the deformation \eqref{pentaladderRecNewRatio} of cross ratios, i.e. for any function $f(u,v,w)$,
\begin{equation}
    \mathcal D[f(\tilde u,\tilde v,\tilde w)]=[\mathcal Df]_{\tilde u,\tilde v,\tilde w}.
\end{equation}

We derive \eqref{pentaladderDiffEq} by induction. It's clear that $\mathcal D\Psi_1(u,v,w)=(1-u-v+uvw)=:\Psi_0(u,v,w)$ as defined in \cite{Drummond:2010cz}. Now, suppose $\mathcal D\Psi_L(u,v,w)=\Psi_{L-1}(u,v,w)$. Acting the differential operator on \eqref{pentaladderRecNew},
\begin{equation}
    \begin{split}
        \mathcal D\Psi_{L+1}(u,v,w) &= \displaystyle\int{\rm d}\log(\tau_Y+1){\rm d}\log\frac{\tau_X+1}{\tau_X}\mathcal D\Psi_L(\tilde u,\tilde v,\tilde w) \\
        &= \displaystyle\int{\rm d}\log(\tau_Y+1){\rm d}\log\frac{\tau_X+1}{\tau_X}[\mathcal D\Psi_L]_{\tilde u,\tilde v,\tilde w} \\
        &= \displaystyle\int{\rm d}\log(\tau_Y+1){\rm d}\log\frac{\tau_X+1}{\tau_X}\Psi_{L-1}(\tilde u,\tilde v,\tilde w) \\
        &= \Psi_L(u,v,w).
    \end{split}
\end{equation}
We remark that it is straightforward to check that WL ${\rm d}\log$ representation for various ladder integrals satisfy similar differential equations. It would be interesting to understand how to obtain such representation for more general integrals by solving differential equations they satisfy.

\subsection{Resumming the ladders}

Now we move to the second question, namely if we can resum certain ladder integrals starting from the fact that  the $L$-loop integral is an integral of $(L{-}1)$-loop one with shifted kinematics, \textit{e.g.} eq.\eqref{penta1}, eq.\eqref{penta2}, eq.\eqref{double1}. We will take the sum over all $L$ for the ladders dressed with coupling constant, which will yield an integral equation for the resummed ladder. It turns out to be possible to solve this integral equation by series expansion, and get a solution for resummed ladder at any value of the coupling. In this subsection, we will discuss in details the resummed penta-ladder $\Psi_g=\sum_{L=1}^\infty g^{2L}\Psi_{L}$, but the same method also works for other ladders, e.g. the double-penta ladder $\Psi_L(1,4,5,7)$ defined in section \ref{sec.4.1}.

Recall that a generic $L$-loop penta-ladder integral $\Psi_L$ only depends on three cross ratios $u$, $v$, $w$, and satisfies the following recursion:
\begin{align*}
    \Psi_{L{+}\frac12}(u,v,w)&=\int_0^\infty {\rm d} \log\frac{t+1}t\ \Psi_{L}\left(\frac{u(t+w)}{t+uw},v,\frac{w(t+1)}{t+w}\right),\\
	\Psi_{L{+}1}(u,v,w)&=\int_0^\infty {\rm d} \log(t+1)\ \Psi_{L{+}\frac12}\left(u,\frac{v(t+1)}{t v+1},\frac{t+w}{t+1}\right).
\end{align*}

We want to solve these equation by series expansion near the fixed point of the transformation of kinematics, $(u=1, v=1,w=1)$ , thus it is convenient to use $\{x=1-1/u,y=1-1/v,z=1-w\}$ as our variables, and define $\Phi_{*}(x,y,z):=\Psi_{*}(u,v,w)$. Then, the recursion now reads
\begin{align}
	\Phi_{L{+}\frac12}(x,y,z)&=\int_0^\infty {\rm d} \log\frac{t+1}t\ \Phi_{L}\left(\frac{tx}{t+1-z},y,\frac{t z}{t+1-z}\right)\\
	\Phi_{L{+}1}(x,y,z)&=\int_0^\infty {\rm d} \log(t+1)\ \Phi_{L{+}\frac12}\left(x,\frac{y}{1+t},\frac{z}{1+t}\right)
\end{align}
with the starting point which is the one-loop chiral pentagon
\begin{align}\label{phi1}
\begin{aligned}
	\Phi_1(x,y,z)=&\log (1-x) \log (1-y)+\Li_2(z)-\Li_2\left(\frac{x-z}{x-1}\right),\\
	&+\Li_2\left(\frac{x}{x-1}\right)-\Li_2\left(\frac{y-z}{y-1}\right)+\Li_2\left(\frac{y}{y-1}\right).
\end{aligned}
\end{align}

Define the resummed version of $\Phi_{L{+}\frac12}$ and $\Phi_{L}$ which depend on the coupling constant $g$:
\[
	\Phi_g^{\text{odd}}:=\sum_{L=1}^\infty g^{2L+1}\Phi_{L{+}\frac12}
	\quad \text{and}\quad 
	\Phi_g^{\text{even}}:=\sum_{L=1}^\infty g^{2L}\Phi_{L}
\]
and then we arrive at integral equations of $\Phi_g^{\text{odd}}$ and $\Phi_g^{\text{even}}$,
\begin{align}
	&\Phi_g^{\text{odd}}(x,y,z)=g\int_0^\infty {\rm d} \log\frac{t+1}t\ \Phi_g^{\text{even}}\left(\frac{tx}{t+1-z},y,\frac{t z}{t+1-z}\right),\\
	&\Phi_g^{\text{even}}(x,y,z)-g^2\Phi_1(x,y,z)=g\int_0^\infty {\rm d} \log(t+1) \Phi_g^{\text{odd}}\left(x,\frac{y}{1+t},\frac{z}{1+t}\right).
\end{align}
Suppose the series expansions of $\Phi_g^{\text{odd}/\text{even}}$ and $\Phi_1$ are
\begin{align*}
\Phi_g^{\text{odd}/\text{even}}(x,y,z)=\sum_{k,l,m=0}^\infty A^{\text{odd}/\text{even}}_{k,l,m}x^k y^l z^m, \quad 
\Phi_1(x,y,z)=\sum_{k,l,m=0}^\infty B_{k,l,m}x^k y^l z^m,
\end{align*}
then we can perform the integrals on the RHS of the integral equations, and obtain recurrence relations for $A^{\text{odd}/\text{even}}_{k,l,m}$:
\begin{align}
	&\sum_{m=0}^\infty A^{\text{odd}}_{k,l,m}z^m=-g\sum_{m=0}^\infty A^{\text{even}}_{k,l,m}\frac{z^m}{k+m}\, _2F_1(1,k+m;k+m+1;z),\\
	&A^{\text{even}}_{k,l,m}-g^2B_{k,l,m}=\frac{g}{m+l}A^{\text{odd}}_{k,l,m}.\label{Arec1}
\end{align}
We first see that $A^{\text{odd}}_{k,0,0}$ and $A^{\text{even}}_{0,l,0}$ must vanish to avoid possible divergences. The first equation can be further simplified by using the identity of the hypergeometric function $_2F_1$
\[
	\frac{d}{dz}(z^{a}\, _2F_1(b,a;a+1;z))=az^{a-1}\, _2F_1(b,a;a;z)=az^{a-1}(1-z)^{-b},
\]
thus it can be simplified significantly (with $A^{\text{odd}}_{k,l,-1}:=0$):
\begin{equation}\label{Arec2}
	(k+m)A^{\text{odd}}_{k,l,m}- (k+m-1)A^{\text{odd}}_{k,l,m-1}=-g A^{\text{even}}_{k,l,m}
\end{equation}
and then we have recurrence relations for $A^{\text{odd}/\text{even}}_{k,l,m}$, eq.\eqref{Arec1} and eq.\eqref{Arec2}. 

It turns out that we can solve these recurrence relations and obtain
\begin{equation}\label{recsolu}
	A^{\text{even}}_{k,l,m}=g^2B_{k,l,m}-g^2\sum_{n=0}^m B_{k,l,n}\frac{g^2}{(k+m)(l+m)}\prod_{s=n}^m\frac{(k+s)(l+s)}{(k+s)(l+s)+g^2}.
\end{equation}
From the explicit expression of $\Phi_1$ eq.\eqref{phi1}, we find
\[
	B_{0,l,0}=B_{k,0,0}=0,\,B_{k,l,0}=\frac{1}{kl}\,\,\text{and}\,\, B_{k,l,m}=\frac{\delta_{k,0}\delta_{l,0}}{m^2}-\frac{\delta_{l,0}}{m(m+k)}-\frac{\delta_{k,0}}{m(m+l)}\,\, \text{for $m\neq 0$},
\]
thus eq.\eqref{recsolu} becomes
\[
	A^{\text{even}}_{k,l,0}=\frac{g^2}{kl+g^2}\,\,\text{for $kl\neq 0$},\,\, A^{\text{even}}_{k,l,m}=-\frac{g^2}{kl+g^2}\frac{g^2}{(k+m)(l+m)}\prod_{n=1}^m\alpha_{k,l,n}\,\,\text{for $m>0$},
\]
where we have defined
\[
\alpha_{k,l,m}:=\frac{(k+m)(l+m)}{(k+m)(l+m)+g^2}.
\]

The upshot is that we obtain the solution of the resummed penta-ladder
\begin{equation}
\boxed{\Phi^{\text{even}}_g(x,y,z)=g^2\sum_{k,l=1}^\infty \frac{x^k y^l }{kl+g^2}-g^2\hspace{-2ex}\sum_{k,l=0,m=1}^\infty\frac{x^ky^lz^m}{kl+g^2}\frac{g^2}{(k+m)(l+m)}\prod_{n=1}^m\alpha_{k,l,n}.}
\end{equation}
This series converges for $|x|,|y|,|z|<1$ and  $g^2\in \mathbb{C}-\{-1,-2,-3,\dots\}$. As a function of $g^2$, it does not have other poles except for negative integers.

One can of course read off $L$-loop penta-ladder integrals from this resummed result, although usually only in the form of series expansion. In some cases, we can have very simple answer, {\it e.g.} when $z=0$ (or $w=1$)
\[
	\Phi^{\text{even}}_g(x,y,0)=g^2\sum_{k,l=1}^\infty \frac{x^k y^l}{kl+g^2}=-\sum_{L=1}^\infty (-g^2)^{L}\sum_{k,l=1}^\infty\frac{x^k y^l}{(kl)^{L}}=-\sum_{L=1}^\infty (-g^2)^{L}\operatorname{Li}_L(x)\operatorname{Li}_L(y),
\]
and the result of $L$-loop integral for $w=1$ is simply $(-1)^{L+1} \operatorname{Li}_L(1-1/u)\operatorname{Li}_L(1-1/v)$.


\section{Conclusion and Discussions}

In this paper we introduced and studied the so-called Wilson-loop ${\rm d} \log$ representation of classes of multi-loop Feynman integrals, which evaluate to generalized polylogarithms of uniform transcendental weight. Generally such representations express a higher-loop integral as ${\rm d}\log$ integrals of lower-loop ones, by converting each loop that satisfies certain conditions into a two-fold ${\rm d} \log$ integral. We have considered one of the simplest families of integrals containing terminal chiral pentagons, where we recursively convert each terminal chiral pentagon into two ${\rm d} \log$'s by using partial Feynman parametrization around massless corners. More concretely we obtain a formula expressing the $L$-loop generalized penta-ladder as $2(L{-}1)$-fold ${\rm d}\log$ integral of a one-loop integral, from which the symbol can be computed easily.

In the most general cases, it can be challenging to evaluate these ${\rm d}\log$ integrals, since one needs to ``rationalize'' possible square roots in the process. However, for simple cases where the ${\rm d}\log$ forms remain linear in the integration variables at all steps, it is straightforward to evaluate such integrals, especially at the symbol level, with our algorithm that bypasses the need to perform any integral. We illustrated how to evaluate the symbol of the simplest examples, {\it i.e.} eight-point penta-ladders and seven-point double-penta-ladders, obtaining their last entries {\it etc.} for arbitrarily high loops without carrying out the integration. Our integral formula for penta-ladders satisfies the familiar differential equation, and quite nicely we can resum the penta-ladders by solving integral equations at any value of the coupling.

The immediate next step is to develop systematic methods for rationalization when evaluating WL integrals involving square roots. We have successfully rationalized all square roots for the most generic ($n=12$) double pentagon integrals, and we expect our results there to generalize. 
In addition, so far we have restricted to ladder-shaped integrals at higher loops, but it would be interesting to find applications in evaluating integrals such as those with the ``ring'' topology and any number of terminal pentagons, as well as those where other one-loop sub-diagrams can be reduced. Even restricted to two loops, can we evaluate all (finite) integrals needed for NMHV and even N${}^2$MHV amplitudes using this method? We have restricted to finite integrals but it would certainly be interesting to study regularization of IR divergences in our representation. 

We have only touched the surface of two potentially very interesting topics calling for further investigations. First, it would be highly desirable to better understand the connections of the WL ${\rm d}\log$ representation and the differential equations they satisfy. For any integral where a loop can be converted into ${\rm d} \log$ forms, one should be able to find a differential operator that ``annihilates'' that loop. Along the other direction, could we find WL ${\rm d}\log$ representations of any integral that satisfies certain differential equations? It would also be interesting to explore connections with differential equations following from Yangian symmetry, such as those in~\cite{Loebbert:2019vcj,Loebbert:2020hxk,Loebbert:2020glj} as well as first-order differential equations \cite{Chicherin:2018rpz,Chicherin:2018ubl}like the ${\bar Q}$ equations for the complete amplitudes, which use the same type of $\tau$ integrals. Furthermore, it would be extremely interesting to study the resummation of other ladder-shaped integrals and even more general cases using our method. 

Finally, we would like to understand the geometric meaning of these ${\rm d}\log$ representation (for usual loop integrands, see~\cite{ArkaniHamed:2012nw,Arkani-Hamed:2013jha, Herrmann:2020oud}). As we have mentioned, as far as the $\tau$ integrations are concerned, all we are doing is integrating canonical ${\rm d}\log$ forms of certain positive geometries (that are not polytopes) over a simplex. It would be interesting to 
understand these positive geometries underlying WL ${\rm d} \log$ forms of certain Feynman integrals. Moreover, just like the Aomoto case, we would like to extract the symbol directly from such geometries, which would take us one step further than our current algorithm, and may provide more insights into the possible geometric meaning of these Feynman integrals and even the integrated amplitudes.

\section*{Acknowledgement}
We would like to thank Nima Arkani-Hamed and Yang Zhang for inspiring discussions, and especially Chi Zhang for collaborations on related projects. This research is supported in part by the Key Research Program of Frontier Sciences of CAS under Grant No. QYZDBSSW-SYS014, Peng Huanwu center under Grant No. 12047503 and National Natural Science Foundation of China under Grant No. 11935013. 
\appendix

\section{Explicit results for the symbol of two-loop integrals}\label{resultsq}
Here we present explicit results of \eqref{O2symbol} and \eqref{Omega2symbol}. Once notation $(a\longleftrightarrow b)$ shows up, one should repeat  all terms before it with $a$ and $b$ exchanged. 
\paragraph{Pentaladder $\Psi_2$}
\begin{equation}
\begin{split}
    \mathcal{S}&(q_w)=u{\otimes} (1{-}u){\otimes}\frac{v (1{-}w) (1{-}u w)}{w (1{-}u{-}v{+}u v w)}{+}u{\otimes} v{\otimes}\frac{w (1{-}u{-}v{+}u v w)}{(1{-}u w) (1{-}v w)}{+}u{\otimes} u{\otimes} \frac{1{-}u w}{1{-}u}\\
    &+(u{\otimes}w{+}w{\otimes}u){\otimes} \frac{v (1{-}w) (1{-}u w)}{(1{-}u) (1{-}v w)}{+}(u w){\otimes} (1{-}u w){\otimes} \frac{(1{-}u) w (1{-}u{-}v{+}u v w)}{v (1{-}w) (1{-}u w)^2}\\
    &+(u\longleftrightarrow v){+}w{\otimes} (1{-}w){\otimes} \frac{u v (1{-}w)^2 (1{-}u w) (1{-}v w)}{(1{-}u) (1{-}v) w (1{-}u{-}v{+}u v w)}
\end{split}
\end{equation}
\begin{equation}
\begin{split}
    \mathcal{S}&(q_{uv})=u{\otimes}(1{-}u){\otimes}\frac{(1{-}u)u v^2 (1{-}w) (1{-}u w)}{(1{-}u{-}v{+}u v w)^2}{+}u{\otimes} u{\otimes} \frac{1{-}u w}{(1{-}u) v}{+}u{\otimes} v{\otimes} \frac{(1{-}u{-}v{+}u v w)^2}{u v (1{-}u w) (1{-}v w)}\\
    &{+}(u{\otimes} w{+}w{\otimes}u){\otimes} \frac{(1{-}w) (1{-}u w)}{1{-}v w}{+}(u w){\otimes} (1{-}u w){\otimes} \frac{(1{-}u{-}v{+}u v w)^2}{u v (1{-}w) (1{-}u w)^2}\\
    &+(u\longleftrightarrow v){+}w{\otimes}(1{-}w){\otimes} \frac{u v (1{-}w)^2 (1{-}u w) (1{-}v w)}{(1{-}u{-}v{+}u v w)^2}
\end{split}
\end{equation}
\begin{equation}
    \begin{split}
   \mathcal{S}&(q_{u/v})=u{\otimes} u{\otimes} \frac{v (1{-}u w)}{1{-}u}{+}u{\otimes} v{\otimes} \frac{u (1{-}v w)}{v (1{-}u w)}{+}(u{\otimes} w{+}w{\otimes}u){\otimes} \frac{(1{-}u w) (1{-}v w)}{1{-}w}\\
   &{+} w{\otimes} (1{-}w){\otimes}v(1{-}u w){+}(u w){\otimes} (1{-}u w){\otimes} \frac{u (1{-}w)}{v (1{-}u w)^2}{+}u{\otimes} (1{-}u){\otimes} \frac{(1{-}u) (1{-}u w)}{u (1{-}w)}\\
   &-(u\longleftrightarrow v)
    \end{split}
\end{equation}

\paragraph{Double-penta ladder $\Omega_2$}

\begin{equation}
    \begin{split}
        \mathcal{S}&(q_{u_1u_2})=u_1{\otimes} (1{-}u_1){\otimes} \frac{(1{-}u_1{-}u_2{+}u_1 u_2 u_4)^2}{u_1 u_2 u_3 (1{-}u_4)}{-}u_4{\otimes} (1{-}u_4){\otimes} \frac{(1{-}u_1)  }{(1{-}u_1{-}u_2{+}u_1 u_2 u_4)}\\
        &{+}(u_1{\otimes} u_3{+}u_3{\otimes} u_1){\otimes} \frac{u_2 (1{-}u_1 u_4)}{1{-}u_1}{+}u_1{\otimes}u_2{\otimes} \frac{(1{-}u_1)(1{-}u_2) u_3}{(1{-}u_1{-}u_2{+}u_1 u_2 u_4)^2}\\
        &{+}(u_1 u_4){\otimes} (1{-}u_1 u_4){\otimes} \frac{(1{-}u_1) u_1 u_3 (1{-}u_4)}{(1{-}u_1 u_4) (1{-}u_1{-}u_2{+}u_1 u_2 u_4)^2}{+}(u_3{\otimes} u_4{+}u_4{\otimes} u_3){\otimes} (1{-}u_1 u_4)\\
        &+(u_1\longleftrightarrow u_2){-}u_3{\otimes} (1{-}u_3){\otimes} u_3{-}u_4{\otimes} (1{-}u_4){\otimes}u_3{-}(u_3{\otimes} u_4{+}u_4{\otimes} u_3){\otimes} (1{-} u_4)
    \end{split}
\end{equation}

\begin{equation}
    \begin{split}
        \mathcal{S}&(q_{u_1/u_2})=u_3{\otimes} (1{-}u_3){\otimes} u_2 (1{-}u_3{-}u_1 u_4){+}(u_2 u_4){\otimes} (1{-}u_2 u_4){\otimes} \frac{1{-}u_2 u_4}{1{-}u_3{-}u_2 u_4}\\
        &{+}((u_2 u_4){\otimes} u_3{+}u_3{\otimes}(u_2 u_4)){\otimes} \frac{1{-}u_3{-}u_2 u_4}{1{-}u_2 u_4}-(u_1\longleftrightarrow u_2)
    \end{split}
\end{equation}

\begin{equation}
    \begin{split}
        \mathcal{S}&(q_{u_4})=(u_1{\otimes} u_3{+}u_3{\otimes} u_1){\otimes} \frac{u_2 (1{-}u_1 u_4)^2}{(1{-}u_1) (1{-}u_3{-}u_1 u_4)}{+}u_1{\otimes} (1{-}u_1){\otimes} \frac{(u_1 u_2 u_4{-}u_1{-}u_2{+}1)^2}{u_1 u_2 u_3 (1{-}u_4)}\\
        &{+}u_1{\otimes} u_2{\otimes} \frac{(1{-}u_1) (1{-}u_2) u_3}{(1{-}u_1{-}u_2{+}u_1 u_2 u_4)^2}{+}(u_1 u_4){\otimes} (1{-}u_1 u_4){\otimes} \frac{u_1 (1{-}u_1) u_3 (1{-}u_4) (1{-}u_3{-}u_1 u_4)}{(1{-}u_1 u_4)^2 (1{-}u_1{-}u_2{+}u_1 u_2 u_4)^2}\\
        &{+}(u_3{\otimes} u_4{+}u_4{\otimes} u_3){\otimes} \frac{(1{-}u_1 u_4)^2}{1{-}u_3{-}u_1 u_4}{+}u_3{\otimes} (1{-}u_3){\otimes} \frac{1{-}u_3{-}u_1 u_4}{u_1 u_4}{-}u_4{\otimes} (1{-}u_4){\otimes} \frac{(1{-}u_1)  }{(1{-}u_1{-}u_2{+}u_1 u_2 u_4)}\\&{+}(u_1\longleftrightarrow u_2)
        {-}u_3{\otimes} (1{-}u_3){\otimes} u_3{-}(u_4{\otimes}u_3{+}u_3{\otimes}u_4){\otimes} (1{-}u_4){-}u_4{\otimes} (1{-}u_4){\otimes}u_3
    \end{split}
\end{equation}

\bibliographystyle{utphys}
\bibliography{main}
\end{document}